\documentclass[intlimits,twoside,a4paper]{article}
\usepackage{amsmath,amssymb}
\usepackage{graphicx}
\usepackage[T2A]{fontenc}
\usepackage[cp1251]{inputenc}
\usepackage[eqsecnum]{cmpj2}

\issue{2013}{16}{2}{23604}

\doinumber{10.5488/CMP.16.23604}

\newcommand{\be}{\begin{equation}}
\newcommand{\ee}{\end{equation}}
\newcommand{\bea}{\begin{eqnarray}}
\newcommand{\eea}{\end{eqnarray}}

\title[Liquid-gas phase transition]%
{Liquid-gas phase transition at and below the critical point}

\author[I.R. Yukhnovskii, V.O. Kolomiets, I.M. Idzyk]{I.R. Yukhnovskii,
        V.O. Kolomiets, I.M. Idzyk}
\address{
Institute for Condensed Matter Physics of the National Academy of Sciences of Ukraine,\\
1 Svientsitskii St., 79011 Lviv, Ukraine
}

\date{Received February 28, 2013, in final form March 30, 2013}

\begin{document}

\maketitle

\begin{abstract}
This article is a continuation of our
previous works~(see Yukhnovskii~I.R. et al., J.~Stat. Phys, 1995, \textbf{80}, 405 and references therein), where we have described the behavior of a simple system
of interacting particles in the region of temperatures at and
about the critical point, $T\geqslant  T_{\mathrm{c}}.$
Now we present a description of the behavior of the system at the
critical point $(T_{\mathrm{c}}, \eta_{\mathrm{c}})$ and in the region below the
critical point. The calculation is carried out from the first
principles. The expression for the grand canonical partition
function is brought to the functional integrals defined on the set
of collective variables. The Ising-like form is singled out. Below $T_{\mathrm{c}}$, when a gas-liquid
system undergoes a phase transition of the first order, i.e.,
boiling, a ``jump'' occurs from the ``extreme'' high probability
gas state to the ``extreme'' high probability liquid state,
releasing or absorbing the latent heat of the transition. The
phase equilibria conditions are also derived.
\pacs 64.70.F-, 64.60.F-, 05.70.Jk
\end{abstract}

\section{Introduction}


In this work we complete the first stage of the study of a system at
the gas-liquid critical point by means of the collective variables
method.

The research in this direction started in the early 1980s with
work~\cite{1}. By that time, the collective variables method had already
been developed in the approach proposed by D. M. Zubarev~\cite{2,5}, as
well as in the Hubbard transformation approach~\cite{3,4}. The application of
this method has been successful with respect to a number of physical
problems in the theory of condensed particle systems interacting via
long-range as well as short-range potentials. An effective
approximate solution to the three-dimensional Ising model was
achieved and applied to describe phase transitions of the second
order in a variety of systems~\cite{7,8,8a,9,9a}. A whole bunch of brilliant
papers and monographs on the phase transition theory has emerged~\cite{10,11,12,13}.

The transformation from the real space of Cartesian coordinates to
the set of collective variables defined in the space of wave vectors
${\bf k}$ provided an obvious advantage in the description of systems of
the interacting particles that attract one another at large
separations. Exactly this attraction, which is usually given by a
long-range ``tail'' of a Van der Waals attraction type, is the
source of liquid-gas phase transitions. In the $k$-space, such
attraction is described by the behaviour of the Fourier transform of
the interaction potential at small $k$'s, and, more importantly, in
the close vicinity of $k=0$. This is one gain of the passing from the
Cartesian space to the wave vector ${\bf k}$ space. Another gain
comes from the set of collective variables. This set contains one
variable (in the gas-liquid system case, it is $\rho_{\bf k}$ for $k=0$)
which is directly linked to the order parameter that characterizes
the phase transition.

Therefore, the system of collective variables (CV) $\rho_{\bf k}$ \cite{2,5} or
their conjugates $\omega_{\bf k}$ \cite{3,4} can be thought of as the
most suitable one for the description of the gas-liquid phase
transition.

The results of the CV method application to a variety of Ising-like
systems, obtained with the precision up to quartic and even sextic
measure density, are presented with a large bibliography in monograph~\cite{14}.

Initial expressions for the partition function, given
here in equations~(\ref{eq1.17}) and~(\ref{eq1.18}), and for the quartic measure density
in equation~(\ref{eq2.2}), were obtained in~\cite{15,16,17,18,19,20,21,22,23}. Similar expressions were
obtained in the works of Hubbard, Hubbard and Scofield~\cite{24}, and
in the work of Vouse and Sac~\cite{25}. In the latter, the contribution from the
transformation Jacobian was counted as addition of some entropic
term. In~\cite{16,17}, the values of the cumulants
$\frak M_n ({\bf k}_1 \dots {\bf k}_n)$ in the vicinity of ${\bf
k}_i =0$ were found. It was shown that for all $\frak{M}_n ({\bf k}_1 \dots {\bf
k}_n)$ at small $k$ near the points $k_i =0$, $i = 1, \dots, n$, there exist plateaus wide
enough for the values of the Fourier transform of the
attractive potential for the regions $k < B$ where $\tilde \Phi(k) < 0$ and $\tilde \Phi(B) = 0$ to lie
entirely within the span of those plateaus of $\frak{M}_n ({\bf k}_1 \dots {\bf
k}_n)$. Moreover, it turns out
that the values $\frak M_n (0 \dots 0)$ can be expressed through the
compressibility of the reference system and its derivatives.

The
nature itself granted us a possibility to bring the problem of the
liquid-gas phase transition to a solvable form.

In the works~\cite{18,19,20}, expressions for the equation of state for $T
\geqslant  T_{\mathrm{c}}$ were obtained. A formula for the critical temperature of a
liquid-gas system was found. The calculations were carried out in
the critical region of temperatures close to $T_{\mathrm{c}}$. The region $T
\leqslant  T_{\mathrm{c}}$ has not received a proper treatment in~\cite{19,20,21,22}. With this
work we renew  the endeavor to
reveal the processes that take place at $T \leqslant  T_{\mathrm{c}}$ in the critical
region. The critical region means a region where a
renormalization-group symmetry characterizes the relations between
the coefficients of the block Hamiltonians.

Before the present work started, a number of authors had
produced a huge amount of immensely interesting works~\cite{26,27,28,29,30,PhysRevE.85.031131,32}.

In the section~2, the starting form of the partition function in
the grand canonical distribution is given in terms of collective
variables $\rho_{\bf k}$. The long-range attraction of a Van der Waals type is described
by the set $\rho_{\bf k}$, whereas the short-range repulsion of an
hard-spheres type is described as a reference system in the phase-space of the Cartesian coordinates of the particles. We
start with a quartic measure density, instead of a Gaussian one. The
curves for the cumulants of the transformation Jacobian are
presented in~\cite{21}. Their form allows us to reduce the problem to the Ising
model in an external field. The role of the latter is played by the
generalized chemical potential $\mu^*$. A lot of brilliant works by M.Kozlovskii were devoted to the research of the Ising model in an external field (see review~\cite{UFZ_09}).

The displacement transformation, applied to the macroscopic
variables $\rho_0$ and $\omega_0$ in order to achieve a proper Ising-like
form, has a profound meaning.

We suppose here that the main events, connected with the phase transition in the vicinity of the critical point, occur in the region $k_i \leqslant   B,$ such that $\tilde{\Phi}(k)<0$ and $\tilde{\Phi}(B)=0.$

Integration in the partition function at $T \leqslant  T_{\mathrm{c}}$ is performed in
three regimes. In the renormalization group regime for the wave
vectors $B_{m_{\tau}} \leqslant  k \leqslant  B,$ the Wilson linear
approximation~\cite{33,34,35,36} in the expansion of the recursive equations
around the fixed point and the Kadanoff's hypothesis of scale
invariance~\cite{37} are used. Further, for $0 < k \leqslant   B_{m_\tau}$, the
integration is carried out in the inverse Gaussian regime (IGR),
just like it is done in the Ising model at $T \leqslant  T_{\mathrm{c}}$ \cite{9}. Let us
note that without integration in the IGR, one would not be able
to obtain the correct behaviour of the system entropy~\cite{9,9a}, even in
the limit $T \to T_{\mathrm{c}}$, $T = T_{\mathrm{c}}$.

And finally, we come to the integration over the variable $\rho_0$. The
corresponding ``Hamiltonian'' would be somewhat an analogue of the Landau
problem~\cite{38}. However, herein everything is done in a
coherent manner and the ``Hamiltonian's'' coefficients are obtained as well as their non-analytic dependence on the temperature.

The most significant part of the work concerns the part of the
partition function,
connected with the generalized chemical potential $\mu^*$ and the
integral over $\rho_0$.
The integration over $\rho_0$ is done by the steepest-descent method. In a way, we
are ``traveling'' along the ridge of the integrand's maxima.

The study of the maxima reveals the behaviour of the generalized chemical
potential $\mu^*$. There was shown the
existence of a region $\mu^* =0$ within which the system experiences
a ``jump'' of the density. It is here that the parameter $\Delta$ appears
when going from the dependence of $P = p(\mu,\tau,\eta)$ to the dependence of $P = p(\tau,\eta)$. The
quantity $\Delta$ is a function of the cumulants that characterize the reference
system. Hereafter, the variable $\rho_0$ is replaced with $\Delta$.
This way, the reference system characterized by the potential
$\psi(r_{ij})$ of hard spheres system [see~(\ref{eq1.5})] ``intrudes''
into the function $E(\rho_0)$ obtained from the integration
corresponding to the long-range attractive potential $\phi(r_{ij})$.

\section{The grand partition function in collective variables representation}
\subsection{Model}

We consider an equilibrium system of interacting equivalent
particles. All its thermodynamical properties are described by the
grand partition function $\Xi$:
\be
\label{eq1.1}
\Xi= \sum\limits_{N=0}^{\infty} \frac{1}{N!}z^N Z_N\,,
\ee
where $N$ is the number of particles, $z$ is the system activity,
\be
\label{eq1.2}
z^N = \left[ \sqrt{\frac{m k_{\rm B}T}{2\pi}}^3 \frac{1}{\hbar^{3}} \right]^N \exp (\beta
\mu N),
\ee
$m$ is the mass of a particle, $k_{\rm B}$ is the Boltzmann
constant, $T$ is the temperature, $\hbar$ is the Planck constant,
$\beta = (k_{\rm B}T)^{-1}$, $\mu$ is chemical potential of the system,
$Z_N$ is the configurational  integral of $N$ particles:
\be
\label{eq1.3}
Z_N = \int \exp ( -\beta \Psi_N)\rd\Gamma_N\,, \ee
$\rd\Gamma_N$ is the
volume element in a phase space of coordinates of particles, $\Psi_N$
is the potential interaction energy. It is equal to a sum of
interactions of two kinds:
\be
\label{eq1.4}
\Psi_N = \frac12 \sum\limits_{{i\leqslant  j, j\leqslant  N}\atop{i\neq j}}
\psi(r_{ij}) + \Phi(r_{ij}), \ee
 where
\be
\label{eq1.5}
\psi(r_{ij}) = \left\{ \begin{array}{cc}
\infty, & r_{ij} \leqslant  \sigma, \\
0, & r_{ij} > \sigma
\end{array}
\right.
\ee
 is the potential interaction energy of two equivalent
hard spheres with a diameter  $\sigma$.

In the present paper we adopt for $\Phi(r_{ij})$ the ``attractive'' branch of the Lennard-Jones potential
\be
\label{eq1.6}
\Psi_{\mathrm{LJ}}(r) = 4\varepsilon \left[ \left( \frac{\sigma_0}{r} \right)^{12} -
\left( \frac{\sigma_0}{r} \right)^{6} \right],
\ee
assuming
\be
\label{eq1.7}
\Phi(r) = \left\{ \begin{array}{cc}
                \Psi_{\mathrm{LJ}}(r), & r \geqslant  \sigma_0\,, \\
                0, & r < \sigma_0\,.
                \end{array}
                \right.
\ee

Values of the parameters $\varepsilon$, $\sigma_0$ in~(\ref{eq1.6}) for Ar, Xe, Kr, O$_2$, CO borrowed from~\cite{41,42} are adduced in table~\ref{tab_a}.

Other functions might be used as $\Phi(r)$ as well. The necessary feature for each of them is the availability of the Fourieur-image
\be
\label{eq1.8}
\tilde \Phi ({\bf k}) = \int\limits_V \Phi (r)\re^{-\ri{\bf k} {\bf r}}\rd{\bf r},
\ee
and condition
\be
\label{eq1.9}
{\rm min}\, \tilde \Phi (k) = \tilde \Phi (0) < 0.
\ee

\begin{figure}[ht]
\centerline{\includegraphics[width=0.95\textwidth]{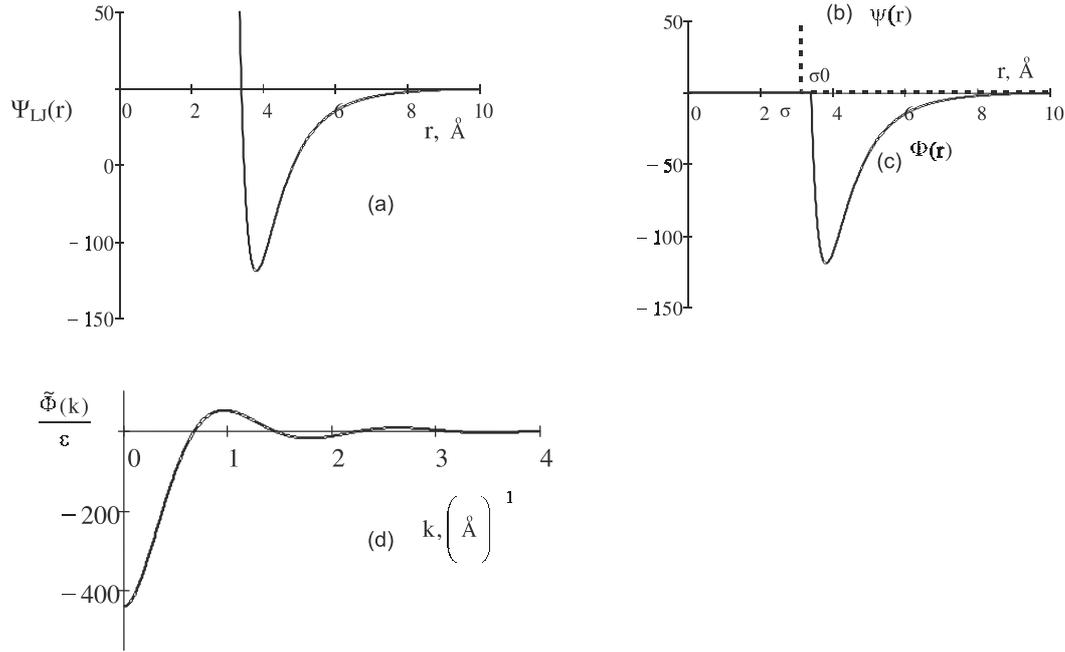}}
\caption{ (a) The full $\Psi_{\mathrm{LJ}}$ potential; (b) the hard spheres potential (for the values of $\sigma$ and $\sigma_0$ see table~\ref{tab_a}); (c) the ``attractive'' long-range potential $\Phi(r)$; (d) the Fourieur-image $\tilde \Phi (k)/\varepsilon$ for the potential $\Phi(r)$.
} \label{potential}
\end{figure}

Plots of $\Psi_{\mathrm{LJ}}(r)$, $\Phi(r)$, $\tilde \Phi ({\bf k}) \equiv \tilde \Phi (k)$ for argon are shown in figure~\ref{potential}.

The phenomena occurring on long scales are of long-wave character. They
are described in  $k$-space by a region of low values of wave
vectors $k$. Thus, we split the space $\{ \bar k \}$ into two subspaces.
Let $B$ correspond to the value $\tilde \Phi(B) =0$, and $\tilde
\Phi(k) <0$ for all $k < B$.

We consider that the main phenomena related to a phase transition occur
in the region $k < B$.

We   pass  on to  an extended phase space which consists of space of
Cartesian coordinates of particles $\{ r \}$ and of space of density
oscillations, collective variables $\{ \rho_k \}$. Overfilling of
the phase space is  eliminated by introducing the ``identity
condition'' in the form of a Jacobian.

\subsection{Reference expressions for the partition function}
\begin{itemize}
\item[a)] {\it Collective variables}. We introduce  the notation:
\be
\label{eq1.10}
\sum\limits_{{i\leqslant  j, j\leqslant  N}\atop{i\neq j}} \Phi(r_{ij}) =
\frac{\langle N \rangle}{V} \sum\limits_{\bf k} \tilde \Phi(k) \bigl[ \hat \rho_N
({\bf k}) \hat \rho_N ({\bf -k}) \bigr] - \frac{\langle N \rangle}{V}\sum\limits_{\bf k} \tilde \Phi(k),
\ee
where $\langle N \rangle$ is the mean number of particles, and we also consider that
\begin{eqnarray*}
\frac{1}{V} \sum\limits_{\bf k} \tilde \Phi(k) \re^{\ri \bf k \bf r}
|_{r \to 0} = \Phi(0) = 0,\\
\hat \rho_N ({\bf k}) = \frac{1}{\sqrt{\langle N \rangle}} \sum\limits_{i=1}^N \exp \left(-\ri \bf k \bf r_i\right)\,.
\end{eqnarray*}

We define the collective
variables system  $\rho_{\bf k}^{\mathrm{c}}$, $\rho_{\bf k}^{\mathrm{s}}$, $\rho_{\bf k}
= \rho_{\bf k}^{\mathrm{c}} - \ri \rho_{\bf k}^{\mathrm{s}}$; $\rho_0$ by the
following relations
\bea
\label{eq1.11}
&& \hat \rho_N^{\mathrm{c}}({\bf k}) = \frac{1}{\sqrt{\langle N \rangle}} \sum\limits_{i=1}^N \cos ({\bf k} {\bf r}_i)= \int \rho_{\bf k}^{\mathrm{c}} J_N ( \rho - \hat \rho_N) (\rd\rho), \nonumber\\
&& \hat \rho_N^{\mathrm{s}}({\bf k}) = \frac{1}{\sqrt{\langle N \rangle}} \sum\limits_{i=1}^N \sin ({\bf k} {\bf r}_i)= \int \rho_{\bf k}^{\mathrm{s}} J_N ( \rho - \hat \rho_N) (\rd\rho), \nonumber\\
&& \hat{\rho}_N(0) = \frac{N}{\sqrt{\langle N \rangle}} = \int \rho_0 J_N (\rho - \hat\rho_N) (\rd\rho).
\eea
Here,
\bea
&& J_N (\rho - \hat \rho_N) = \delta [\rho_0 - \hat\rho_N(0)] \sum_{\bf k}{\vphantom{\sum\nolimits}}' \delta \bigl[ \rho_{\bf k}^{\mathrm{c}} - \hat \rho_N^{\mathrm{c}}({\bf k}) \bigr] \delta \bigl[ \rho_{\bf k}^{\mathrm{s}} - \hat \rho_N^{\mathrm{s}}({\bf k}) \bigr], \nonumber\\
&& (\rd\rho) = \rd\rho_0 \prod_{\bf k}{\vphantom{\sum\nolimits}}' \rd\rho_{\bf k}^{\mathrm{c}} \rd\rho_{\bf k}^{\mathrm{s}}\,.
\nonumber
\eea
Prime means restriction of ${\bf k}$ only to the
values from the upper subspace.
\item[b)] Our {\it reference system} is a system of hard spheres with
diameter $\sigma$, and interaction potential, defined by equation~(\ref{eq1.5}), with chemical potential $\mu_0$ and partition function $\Xi_0$,
%
\be
\label{eq1.12}
\Xi_0 = \sum\limits_{N=0}^\infty \frac{1}{N!} z_0^N \exp(\beta \mu_0 N) \int \exp
[-\beta \psi_N (r)] \rd\Gamma_N\,,
\ee
where $\psi_N(r) = \frac12
\sum\limits_{ij} \psi(r_{ij})$, $z_0^N = (\!\!\sqrt{2m \pi k_{\rm B}
T}/2\pi \hbar)^{3N}$, $\rd\Gamma_N=\rd\textbf{r}_1\rd\textbf{r}_2 \dots \rd\textbf{r}_N$, $\mu_0$ is the reference system chemical potential.
\item[c)] {\it Expression for the partition function in an extended phase
space}.

According to the definitions~(\ref{eq1.1}), (\ref{eq1.11}) and (\ref{eq1.12}):
\bea
\label{eq1.13}
\Xi = \Xi_0 \sum\limits_{N=0}^\infty \frac{z_0^N}{N!} \exp(\beta\mu_0 N) \iint \frac{\exp (- \beta \psi_N)}{\Xi_0} J_N (\rho - {\hat\rho}_N) %
 \exp \left[ h \sqrt{N} \rho_0 - \frac12 \sum\limits_{\bf k}
\alpha (k) \rho_{\bf k} \rho_{-{\bf k}} \right]  (\rd\rho) \rd\Gamma_N\,,\!\!\!\!\!\!\!\!\!\!\!\!\!\!\!\!\nonumber\\
\eea
where
 \be
 h = \beta (\mu -
\mu_0); \qquad \alpha(k) = \frac{\langle N \rangle}{V}\beta \tilde \Phi(k); \qquad  \Phi(r) =
\frac{1}{V} \sum\limits_{\bf k} \tilde \Phi(k) \re^{-\ri{\bf kr}}.
\nonumber \ee

We introduce the Jacobian function
\be
\label{Jac1}
J(\rho) = \sum\limits_{N=0}^\infty \frac{z_0^N}{N!} \exp(\beta\mu_0 N) \int
\frac{1}{\Xi_0} \exp (- \beta \psi_N) J_N (\rho - \hat \rho_N)
\rd\Gamma_N\,. \ee
After its substitution into (\ref{eq1.13}) we obtain:
\be
\label{eq1.15}
\Xi = \Xi_0 \int \exp \left\{ \sqrt{\langle N \rangle }h \rho_0 - \frac12
\sum\limits_{\bf k}  \alpha(k) \rho_{\bf k} \rho_{-{\bf k}}
 \right\} J(\rho) (\rd\rho). \ee
\end{itemize}

\subsection{Jacobian}

Instead of Dirac delta-functions of expression $J_N(\rho -
\hat\rho_N)$ given by~(\ref{eq1.11})  we use their integral
representation of the type
\[
\delta\bigl[ \rho_{\bf k}^{\mathrm{c}} - \hat \rho_N^{\mathrm{c}}({\bf k})\bigr] =
\int\limits_{-\infty}^{\infty} \exp \bigl\{ \ri 2\pi \bigl[ \rho_{\bf
k}^{\mathrm{c}} - \hat \rho_N^{\mathrm{c}}({\bf k})\bigr] \omega_{\bf k}^{\mathrm{c}} \bigr\} \rd
\omega_{\bf k}^{\mathrm{c}}\,.
\]
Then,
\be
\label{eq1.16}
J(\rho) = \int \exp \Bigl(\ri 2\pi \sum\limits_{\bf k} \omega_{\bf
k} \rho_{\bf k}\Bigr) \tilde J (\omega) (\rd\omega),
\ee
where
$\rho_{\bf k} = \rho_{\bf k}^{\mathrm{c}} - \ri \rho_{\bf k}^{\mathrm{s}}$, $\omega_{\bf k}
= \frac12 (\omega_{\bf k}^{\mathrm{c}} + \ri \omega_{\bf k}^{\mathrm{s}})$, $(\rd\omega) = \rd\omega_0
\prod\limits_{\bf k}{\vphantom{\sum\nolimits}}' \rd\omega_{\bf k}^{\mathrm{c}} \rd\omega_{\bf k}^{\mathrm{s}}$, $\tilde
J (\omega)$ is ``the Fourier transform'' of $J(\rho)$,
\be
\tilde J(\omega) = \sum\limits_{N=0}^\infty \frac{z_0^N}{N!} \exp(\beta \mu_0N) \Xi_0^{-1}
\int \exp (-\beta \psi_N) \exp \bigl[- \ri2\pi \sum\limits_{\bf k}
\omega_{\bf k} \hat \rho_N ({\bf k}) \bigr] \rd\Gamma_N\,. \nonumber
\ee

After integration and summation we get $\tilde
J(\omega)$ in an exponential form:
\bea
\label{eq1.17}
\tilde J(\omega) &=&  \exp \left\{ -\ri 2\pi {\frak M}_1 \frac{1}{\sqrt{\langle N \rangle}}\omega_0 - \frac{(2\pi)^2}{2} \frac{1}{\langle N \rangle} \sum\limits_{\bf k} {\frak M}_2 (k) \omega_{\bf k} \omega_{-{\bf k}} \right\}  \nonumber\\
&&{} \times \exp \left\{ \sum\limits_{m \geqslant  3} \frac{(-i2\pi)^m}{m!}
\frac{1}{\sqrt{\langle N \rangle^m}}\sum\limits_{{\bf k}_1, \dots, {\bf k}_m} {\frak M}_m ({\bf k}_1,
\dots, {\bf k}_m) \omega_{{\bf k}_1} \dots \omega_{{\bf k}_m}
\right\}.
\eea
Here, ${\frak M}_1, {\frak M}_2, \dots, {\frak M}_m$
are cumulants of the reference system.

We substitute the expression for  $J(\omega)$ into~(\ref{eq1.16}). Then, we
substitute the obtained result into~(\ref{eq1.15})   and get
\begin{eqnarray}
\label{eq1.18}
    \Xi &=& \Xi_0 \int \exp \left[ \sqrt{\langle N \rangle} h \rho_0 - \frac12
    \sum\limits_{\bf k} \alpha (k) \rho_{\bf k}\rho_{-{\bf k}} \right]%
    \exp \left(\ri2\pi \sum\limits_{\bf k} \omega_{\bf k} \rho_{\bf k} \right)
    \tilde J(\omega) (\rd\rho) (\rd\omega).
\end{eqnarray}
 All expressions entering~(\ref{eq1.18}), $h$, $\alpha (k)$ and cumulants ${\frak M}_n$ are the
functions of density, temperature. This is  the starting formula for
the study of the grand partition function.

\section{Integration of the grand partition function: the phase transition investigation at and below $T_{\mathrm{c}}$}

\subsection{Separation of an integration region in  $k$ space}

\begin{figure}[ht]
\centerline{\includegraphics[angle=0,width=0.85\textwidth]{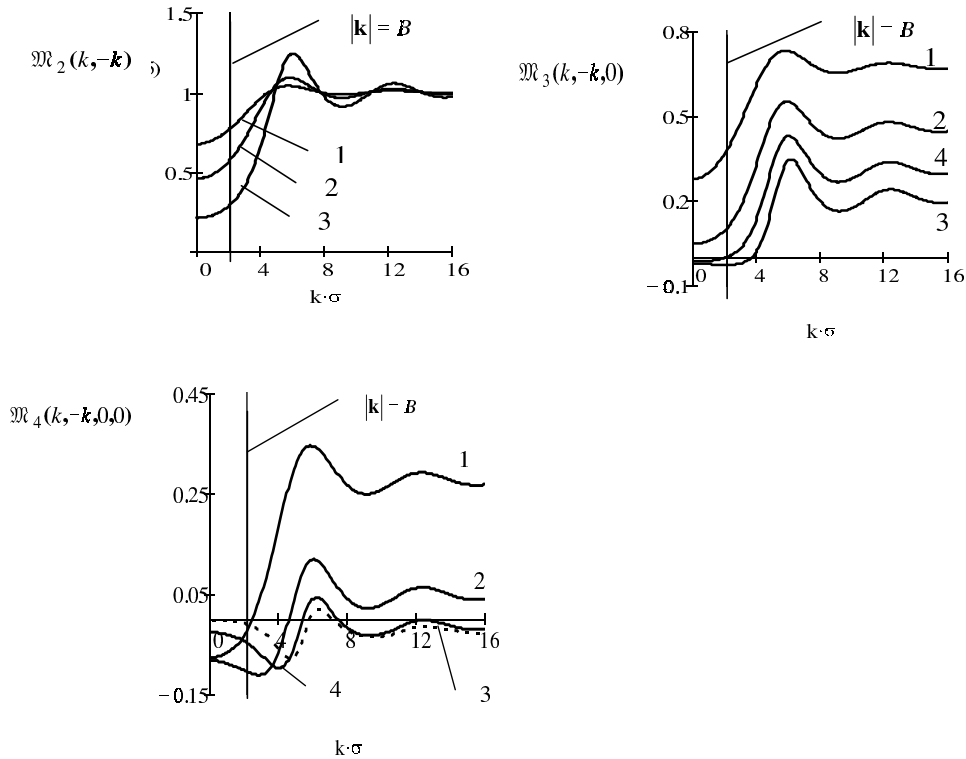}}
\caption{Curves of cumulants $\frak{M_2}(k,-k)$, $\frak{M_3}(k,-k,0)$, $\frak{M_4}(k,-k,0,0)$ for different densities: 1~--~$\eta=0.05,$ 2~--~$\eta=0.1,$ 3~--~$\eta=0.2,$ 4~--~$\eta=0.15.$ Vertical line indicates the value $|{\bf k}| = B$ for which $\Phi(B) = 0$; $\eta = \frac{N}{V}\frac{\pi}{6}\sigma^3$.
} \label{cumulant}
\end{figure}

Let us compare the form of the curves of $\tilde \Phi (k)$ with
that of ${\frak M}_2(k)$. As it follows from  figure~\ref{potential} and as it was
agreed upon, $\tilde \Phi (k)$ at $k=0$ is a negative and finite
quantity, $\tilde \Phi (k)$ is going to zero with an increasing
$k$ and at  $k \to \infty$.
Curve of  ${\frak M}_2(k)$ is given in figure~\ref{cumulant}.

We suppose that the main attraction effects created by potential $\Phi
(r)$ are concentrated in the expression for $\tilde \Phi (k)$ in the
narrow region of $k$ between the values $0$ and $B$. For these values of
$k,$ the curves for  cumulant ${\frak M}_2(k)$ and for all cumulants
${\frak M}_n(k)$ have wide plateaus (see figure~\ref{cumulant}) that begin  for ${\bf k}_1,
\dots, {\bf k}_n = 0$~\cite{21}.  Thus, for all cumulants ${\frak M}_n({\bf
k}_1 \dots {\bf k}_n)$ in the region ${\bf k}_i < B$ we are able to
choose their values for  ${\bf k}_i =0$ (see table~\ref{cum_tab}). This means that
\[
{\frak M}_n({\bf k}_1, \dots, {\bf k}_n) = {\frak M}_n(0, \dots,
0)\quad {\mbox {for }\quad  {\bf k}_i < B}.
\]

\begin{table}[h]
\caption{The reference system cumulants at zero values of the wave vectors and the coefficients $a_2$, $a_4$ for argon at different densities $\eta$ \cite{19}.}
\vspace{2ex}
\label{cum_tab}
\begin{center}
\begin{tabular}{|c|c|c|c|c|c|}
\hline
  $\eta$ & ${\frak M_2}(0,\eta)$ & ${\frak M_3}(0,\eta)$ & ${\frak M_4}(0,\eta)$ & $a_2$ & $a_4$ \\
\hline\hline
0.08 & 0.589 & 0.108  & --0.216 & 1.38 & 0.53  \\
0.10 & 0.471 & 0.048  & --0.137 & 1.73 & 0.82\\
0.12 & 0.329 & 0.012  & --0.079 & 2.13 & 1.13\\
0.14 & 0.337 & --0.008 & --0.043 & 2.57 & 1.39\\
0.16 & 0.296 & --0.019 & --0.022 & 3.05 & 1.50\\
0.18 & 0.272 & --0.023 & --0.010 & 3.47 & 1.24\\
0.20 & 0.277 & --0.024 & --0.004 & 3.53 & 0.54\\
\hline
\end{tabular}
\end{center}
\end{table}

These quantities are macroscopic. Their values are equal to
the corresponding  fluctuations in the number of particles of  the
reference system.

We have an equation
\[
{\frak M}_n(0 \dots 0) = \frac{\partial^n \ln \Xi_0}{\partial (\beta
\mu_0)^n} = \frac{\partial^{n-1} \langle N \rangle_0}{\partial
(\beta \mu_0)^{n-1}}\,.
\]
Therefore,
\bea
&& {\frak M}_1(0) = \langle N \rangle_0\,; \qquad {\frak M}_2(0) = \frac{1}{\langle N\rangle} \left\langle \Bigl(N - \langle N \rangle \Bigr)^2 \right\rangle_0; \nonumber\\
&& {\frak M}_3(0) = \left\langle \Bigl(N - \langle \langle N \rangle \rangle \Bigr)^3
\right\rangle_0 \frac{1}{\sqrt{N}^3}\,;
 \qquad {\frak M}_4(0) = \frac{\left\langle
\Bigl(N - \langle N \rangle \Bigr)^4 \right\rangle_0 - 3 \left\langle \Bigl(N
- \langle N \rangle \Bigr)^2 \right\rangle^2_0}{\langle N\rangle^2}\,, \nonumber
\eea
and so on.

The cumulants ${\frak M}_n(0 \dots 0)$ are functions of the chemical
potential $\mu_0$ and of the density.  One can perform
the elimination of the dependence on $\mu_0$ either in the standard way
extracting the value of $\mu_0$ from the equation $\frac{\partial \ln
\Xi_0}{\partial \mu_0} = \langle N \rangle$, or using for fluctuations $\langle (N -
\langle N \rangle )^n \rangle$ their values for canonical ensemble.
\[
\left( \frac{\partial \langle N \rangle}{\partial \mu_0}
\right)_{TV} = \frac{N}{V} {\kappa}.
\]
We suppose that $\langle N \rangle_0$ coincides with $\langle N \rangle$ and we put here and then $N \equiv \langle N \rangle = \langle N \rangle_0\,$.
Then,
\bea
&& {\frak M}_2(0) = Nk_{\rm B}T\frac{1}{v}{\kappa}, \qquad {\frak M}_3(0) = N(k_{\rm B}T)^2 \left[ 2 \left( \frac{1}{v}{\kappa} \right)^2 - \frac{{\kappa}}{v} \frac{\partial {\kappa}}{\partial v} \right], \nonumber\\
&& {\frak M}_4(0) = N(k_{\rm B}T)^3 \frac{{\kappa}}{v} \left[ 6 \left(
\frac{{\kappa}}{v} \right)^2 - 6 \left( \frac{{\kappa}}{v}
\frac{\partial {\kappa}}{\partial v} \right) + \left( \frac{\partial
{\kappa}}{\partial v} \right)^2 + {\kappa} \frac{\partial^2
{\kappa}}{\partial v^2} \right], \nonumber
\eea
where $v =
\frac{V}{N}$, ${\kappa} = - \frac{1}{V} \left( \frac{\partial
V}{\partial P} \right)_{TN}$ is compressibility in the reference
system. Here, we should refer to an equivalency of the results for
compressibility and its derivatives obtained for canonic and grand
canonical ensembles.

As concerns the dependence of the cumulants ${\frak M}_n(k_1, \dots,
k_n)$ on $k_i$, as it was shown in~\cite{15}, the following expansion is valid
\[
{\frak M}_n(k_1, \dots, k_n) = {\frak M}_n(0, \dots, 0) + C_n^2
{\frak M}_{n-2}(0 \dots 0) \mu_2 (k) k^2 + \dots ,
\]
where $\mu_2 (k)$ is the pair correlation function of the reference
system and $C_n = \frac{n(n-1)}{2}\,$.

The above described situation for the cumulant  values for $k_i < B$
and for $k=0$
 has become a real key to the solution of a problem of the
liquid-gas critical point as well as to the description of
the phenomena related to a liquid-gas phase transition, that is to the boiling
processes occurring at temperatures below $T = T_{\mathrm{c}}$. The graphs for the cumulants ${\frak M}_n(k_1 \dots k_n)$ were obtained in~\cite{16,17}.

As concerns the integration of the mixed terms, in particular, the integration of the
expression
\[
\frac{(2\pi \ri)^4}{2} \sum\limits_{{k_1 >
B}\atop{k<B}}{\frak M}_4 ({\bf k}_1, - {\bf k}_1, {\bf k}, -{\bf k})\omega_{{\bf
k}_1}\omega_{-{\bf k}_1}\omega_{\bf k}\omega_{-{\bf k}}\,,
\]
the
correction to $\sum\limits_{k<B}{\frak M}_2(0,0)\omega_{{\bf
k}}\omega_{-{\bf k}}$ is  less than one percent of the value ${\frak
M}_2(0,0)$.

As a result, we have the following  reference expression for $\Xi$:
\be
\label{eq2.1}
\Xi = \Xi_0 \Xi_{\mathrm{G}} \Xi_{\mathrm{L}}\,,
 \ee
where $\Xi_0$ is the partition
function of the reference system, $\Xi_{\mathrm{G}}$ is the result of
integration over $\rho_{\bf k}$ and over $\omega_{\bf k}$ for the values ${\bf
k}>B$. We also suppose that the integration over $\rho_{\bf k}$ for ${\bf k} >B$ can be fulfilled in the well-known way. The expression for $\Xi_{\mathrm{G}}$ with the accuracy up to the fourth virial coefficient is presented in~\cite{21} in Appendix~A. Note, that this quantity does not affect the critical behavior.

The quantity $\Xi_{\mathrm{L}}$ is the partition function in the region $k<B$.
\be
\label{eq2.2}
\Xi_{\mathrm{L}} = \int w_4 (\rho\omega) (\rd\rho)^{N_B}(\rd\omega)^{N_B}\,. \ee

Integrals over $\rho_k$ and $\omega_k$ in the region  $k <
B$ are taken with quartic basic measure density\footnote{To
be correct, with quartic measure density for integration over
$\omega_k$, if ${\frak M}_4(0\dots 0) < 0$ and with sextic measure
density, if ${\frak M}_4(0)$ > 0 and $(i)^6M_6<0$.}:
\bea
\label{eq2.3}
w_4(\omega\rho) &=& \exp \left\{ %
\vphantom{\sum\limits_{{k_1 \dots k_n}\atop{k_i < B}}}
h \sqrt{N}\rho_0
    - \frac12  \sum\limits_{k<B} \alpha(k) \rho_{\bf k} \rho_{-{\bf k}}
    + \ri2\pi \sum\limits_{k<B} \omega_{\bf k} \rho_{\bf k}  \right.  \nonumber\\
&&{} \left.+
\sum\limits_{n=1}^4 \frac{(-\ri 2\pi)^n}{n!} N^{1 - n/2} {\frak M}_n
\sum\limits_{{k_1 \dots k_n}\atop{k_i < B}} \omega_{{\bf k}_1} \dots
\omega_{{\bf k}_n} \delta_{{\bf k}_1 + \dots + {\bf k}_n} \right\}.
\eea
We want to exclude the cubic term from the expression in the exponent function~(\ref{eq2.3}).
This can be performed by two substitutions
\be
\label{eq2.4}
\omega_0 = \omega'_0 + \frac{\sqrt{N} {\frak M}_3}{(2\pi \ri)
{\frak M}_4}~ \mbox{and}~~ \rho_0 = \rho'_0 + \tilde {\frak M}_1\,,
\ee
 where
\[
 \tilde {\frak M}_1 = \sqrt{N} \Bigl( 1 + {\frak M}_2\xi + \frac13{\frak M}_3\xi^2 \Bigr), \qquad
 \xi = \frac{{\frak M}_3}{|{\frak M}_4|}\,.
\]
From now
on, we omit the argument  $(0)$ in the notations of cumulants and
write down ${\frak M}_2(0) \equiv {\frak M}_2$, ${\frak M}_3(0) \equiv
{\frak M}_3$, ${\frak M}_4(0) \equiv {\frak M}_4,$ etc. (see table~\ref{cum_tab}). After some
tedious transformations we get  $\Xi_{\mathrm{L}}$ in the following form\footnote{Primes at $\omega'_0$ and $\rho'_0$ are omitted.}:
\bea
\label{eq2.5}
 \Xi_{\mathrm{L}} &=& {\bf \Upsilon} \int \exp \left\{ \sqrt{N} \mu^* (\rho_0 + \tilde {\frak M}_1 ) - \frac12 \sum\limits_{k<B} \alpha(k) \rho_{\bf k} \rho_{-{\bf k}} +  \right.%
\ri 2\pi \sum\limits_{k<B} \omega_{\bf k} \rho_{\bf k} -
\frac12 (2\pi)^2 \sum\limits_{k<B} \tilde {\frak M}_2
\omega_{\bf k} \omega_{-{\bf k}}
\nonumber\\
&&{} \left. - \frac{(2\pi)^4}{4!} \frac{1}{N_B}
\sum\limits_{k<B} | \tilde {\frak M}_4 | \omega_{{\bf k}_1} \dots
\omega_{{\bf k}_4} \delta_{{\bf k}_1 + \dots + {\bf k}_4} \right\}
(\rd\omega)^{N_B} (\rd\rho)^{N_B},
\quad{}
\eea
where ${\bf \Upsilon} = {\bf \Upsilon}_0 \exp \left\{ \frac12 |\alpha(0)|\tilde {\frak M_1^2} \right\}$,
\[
{\bf \Upsilon_0} = \exp \left\{ N \Bigl(\xi +\frac12{\frak{M}}_2\xi^2 + \frac{1}{3}{\frak{M}}_3\xi^3+\frac{1}{4!}{\frak{M}}_4\xi^4\Bigr) \right\},
\]
\bea
&& \mu^* = h - \xi + |\alpha(0)| \frac{\tilde {\frak M}_1}{\sqrt{N}}\,, \nonumber\\
&& \tilde {\frak M}_2 = {\frak M}_2 + \frac12 \frac{{\frak M}_3^2}{|\frak M_4|}, \qquad \tilde {\frak M}_4 = \frac{N_B}{N}{\frak M}_4\,, \nonumber\\
&& \tilde {\frak M}_1 = \sqrt{N} [1 - \Delta], \qquad \xi =
\frac{{\frak M}_3}{|{\frak M}_4|}\,, \qquad \Delta = - \Bigl({\frak
M}_2 \xi + \frac{1}{3!} {\frak M}_3 \xi^2\Bigr). \nonumber
\eea
We have obtained the first fundamental result.
We  bring the expression  for  $\Xi_{\mathrm{L}}$ into the form which is
analogous to the form of the partition function of three-dimensional
Ising model in a field of generalized chemical potential  $\mu^*$.
This means that we have built a mathematical framework for the study
of a phase transition.

It seems useful for every physical system that undergoes the phase transition to introduce some analogue of a crystal lattice, on which this transition can be effectively described. Therefore, we treat the quantity  $B$ as a border of  the Brillouin zone for a
simple cubic lattice with the spacing  $c = \frac{\pi}{B}$. The
number of lattice sites in a volume $V$ is equal to $N_B =
\frac{V}{c^3} = V \left( \frac{B}{\pi} \right)^3$, $V =
\frac{\pi}{6}\sigma^3 \frac{N}{\eta}$.

In the case of potential~(\ref{eq1.4}), we have
\be
\label{eq2.6}
1 \leqslant  \frac{N_B}{N} \leqslant  4, \quad {\mbox {\rm~when}}\quad  0.06 \leqslant  \eta \leqslant  0.22; \qquad N_B = N \frac{(B\sigma)^3}{6\pi^2\eta}\,.
\ee

Two essential points should be emphasized in our formulations:
\begin{itemize}
\item[1)] The existence of plateaus for cumulant values  ${\frak
M}_n(k_1 \dots k_n)$ in the region of negative  values of
Fourier-image of attraction potential;
\item[2)] The possibility to introduce the crystal lattice in order to study
the problem of a liquid-gas critical point and to reduce this
problem to the Ising model in an external field.
\end{itemize}

Integration in expression~(\ref{eq2.5}) is taken over $\omega_{\bf k}$. Since
the coefficients  ${\frak M}_n$ do not depend on $k$, passing from
$\omega_{\bf k}$ to $\tilde \omega_{\bf l}$
\[
\tilde \omega_{\bf k} = \sum\limits_{l=1}^{N_B} \tilde \omega_{\bf
l} \re^{-\ri{\bf kl}}
\]
and replacing
\[
\delta_{{\bf k}_1 + \dots + {\bf k}_4} = \frac{1}{N_B} \sum\limits_l
\re^{\ri({\bf k}_1 + \dots + {\bf k}_4)l}
\]
we factorize the integrals over $\tilde \omega_{\bf l}$ in~(\ref{eq2.5})  and
get a reference expression for integration over $\rho_{\bf k}$~\cite{9}
\be
\label{eq2.7}
\Xi_{\mathrm{L}} = \Bigl(Z (\tilde {\frak M}_2 \tilde {\frak M}_4)\Bigr)^{N_B}{\bf
\Upsilon} \Xi_{\mathrm{L}}^{(1)},
\ee
where
\bea
\label{eq2.8}
&& \Xi_{\mathrm{L}}^{(1)} = \int \exp \left\{ \sqrt{N} \mu^* (\rho_0 + \tilde {\frak M}_1) - \frac12  \sum\limits_{k<B} d_2 (k) \rho_{\bf k} \rho_{-{\bf k}} - \frac{a_4}{4!N_B} \sum \rho_{{\bf k}_1} \dots \rho_{{\bf k}_4} \delta_{{\bf k}_1 + \dots +{\bf k}_4} \right\} (\rd\rho)^{N_B},
\eea
\bea
\label{eq2.9}
&& d_2(k) = a_2 + \alpha(k), \qquad \alpha(k) = \frac{N}{V}\beta \tilde \Phi(k), \nonumber\\
&& a_2 = \left( \frac{3}{|\tilde{\frak{M}}_4|} \right)^{-\frac12}U(y), \quad
a_4 = \frac{3}{|\tilde{\frak{M}}_4|} \varphi(y), \qquad y = \sqrt{3} \tilde{\frak{M}}_2 |\tilde{\frak{M}}_4|^{-\frac12}, \nonumber\\
&& U(y) = \frac{U(1,y)}{U(0,y)} > 0, \qquad  \varphi(y) = 3U^2(y) + 2yU(y) - 2 > 0,
\eea
$U(a,y)$ is the Veber parabolic cylinder function for order $a$ and argument~$y$,
\bea
 U(a,y) &=& \frac{2}{\Gamma \Bigl( a + \frac12 \Bigr)}\re^{-\frac{y^2}{4}} \int\limits_0^\infty \exp \Bigl( -yt^2 - \frac12 t^4\Bigr)\rd t, \nonumber\\
 \Bigl(Z({\frak M}_2, {\frak M}_4)\Bigr)^{N_b} \!\!\!&=& 2^{N_b} \sqrt{2}^{N_b-1}
\left[ (2\pi)^{-\frac12} \left( \frac{3}{|\frak{M_4}|} \right)^{\frac14} \exp \left( \frac{y^2}{4} \right) U(0,y) \right]^{N_b}. \nonumber
\eea

It is important to have $d_2(B)>0$ and $d_2(0)<0$. We are working in the narrow temperature region containing the critical point $T=T_{\mathrm{c}}.$

The integral~(\ref{eq2.8}) describes the phenomena in the critical point $T = T_{\mathrm{c}}$,
as well as in the critical region $T
> T_{\mathrm{c}}$ and $T < T_{\mathrm{c}}.$ $|T-T_{\mathrm{c}}|\leqslant  0.01 T_{\mathrm{c}}.$

After integrating over $\rho_{\bf k}$, {\it apart from integration
over} $\rho_0$, the form of the integral~(\ref{eq2.8}) completely coincides
with the corresponding expression for the Ising model. For the purpose of its
integration, we use Kadanoff's idea concerning the scale invariance of
the phenomena on block lattices as well as Wisons's idea concerning the use of a linear
approximation for the expansions around the fixed point in recurrent
relations~\cite{33,34,35,36}.

\begin{figure}[ht]
\centerline{\includegraphics[angle=0,width=0.95\textwidth]{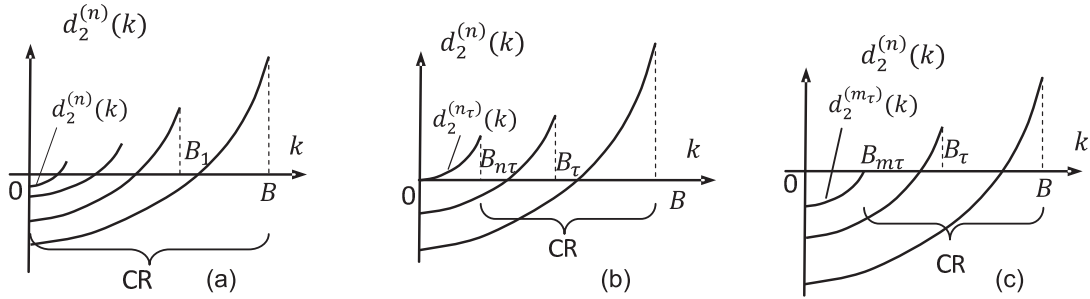}}
\caption{Step-wise integration of the partition function $\Xi_{\mathrm{L}}$ on the collective variables.  (a) $T=T_{\mathrm{c}}:$ $d_2^{(j)}(0)<0$ and $d_2^{(j)}(B_j)>0$ after integration on every layer of the $\rho_{\bf k}$ space, $B_{j+1}<k\leqslant  B_j,$ $j=1,2,...$ (the critical regime, CR);
(b) $T>T_{\mathrm{c}}:$ after integration on variables $\rho_{\bf k},$ $B_{n_{\tau}}<k\leqslant  B$ (the critical regime interval), coefficient $d_2^{(n_{\tau})}(k)$ is positive at $0\leqslant  k \leqslant  B_{n_{\tau}}$ (the limiting Gaussian regime); (c) $T<T_{\mathrm{c}}:$ after integration on variables $\rho_{\bf k},$ $B_{m_{\tau}}<k\leqslant  B$ (the critical regime interval), here $d_2^{(m_{\tau})}(k)<0$ at $0\leqslant  k \leqslant  B_{m_{\tau}}$ (the inverse Gaussian regime).} \label{integration}
\end{figure}

Integration is performed in the real three-dimensional space without
any \emph{a priori} statements about the temperature dependence of the appearing coefficients. It is described in more detail in~\cite{7,8,8a,9,9a} and~\cite{14,16,17,18,19,20,21}. Performing step-wise integration in~(\ref{eq2.8}) on the layers of $\rho_{\bf k},$ $|k|\in (B_1,B], \ldots , |k|\in (B_{n+1},B_n], \ldots$ produces a sequence of effective block Hamiltonians with different types of evolution at $T>T_{\mathrm{c}}$ and $T<T_{\mathrm{c}}$ of coefficients $d_2^{(n)}(k)$ and $a_4^{(n)}$ (see figure~\ref{integration}).

Resuming these results we come to the following expression for the partition function:
\bea
\label{xi_prod}
&& \Xi = \Xi_0\Xi_{\mathrm{G}} \Xi_{\mathrm{L}}^{(\rho_0)} \Xi_{\rho_0}\,.
\eea
Here, $\Xi_0$ is a partition function of the reference system; $\Xi_{\mathrm{G}}$ is the result of integrating over variables $\rho_k$, $\omega_k$ for $k>B$; $\Xi_{\mathrm{L}}^{(\rho_0)}$ is the result of integration in~(\ref{eq2.8}), not including the integration over $\rho_0$\,;
\begin{equation}
\label{xi0}
\Xi_{\rho_0}=\exp\left[\sqrt{N}\mu^*\tilde{\frak{M}}_1\right]\int \exp \left[\sqrt{N}\mu^*\rho_0+D\rho_0^2-\frac1N G\rho_0^4\right] {\rm d}\rho_0\,.
\end{equation}
Coefficients $D$ and $G$ arise as a result of the integration in~(\ref{xi_prod}) over $\rho_{\bf k}.$ This integration was fulfilled in two different ways: with quartic density measure  for $k$ in the interval $B_{m_{\tau}}\leqslant  k\leqslant  B,$ where the renormalization group symmetry is valid, and with Gaussian density measure for $k$ in the interval $0<k\leqslant  B_{m_{\tau}}.$ Here, $B_{m_{\tau}}=B/s^{m_{\tau}},$ $s$ being the parameter of dividing the phase space into layers. The most convenient value for $s$ is $s=s^*=3.58.$ In such a case, we have the values for $m_{\tau}$ presented in table~\ref{mtau}, and coefficients $D$ and $G$ are
$D=D_0|\tau|^{2\nu}$, $G=G_0|\tau|^{\nu}$ with $D_0=1.19$ and $G_0=1.67,$ $\nu=\ln s^* / \ln E_1 = 0.605,$ where $E_1$ is the greater of the two eigenvalues for the matrix of linear recursion equations. Note, that $s<E_1 < s^2,$ and $s>1.$
\begin{table}[h]
\caption{Values of the quantity $m_\tau$.}
\label{mtau}
\begin{center}
\begin{tabular}{|c||c|c|c|c|c|}
\hline
$\tau$ & 0.01 & $10^{-3}$ & $10^{-4}$ & $10^{-6}$ & $10^{-10}$ \\
\hline
$m_{\tau}$ & 1.83 & 2.91 & 4.01 & 6.19 & 10.55 \\
\hline
\end{tabular}
\end{center}
\end{table}

We assume that in~(\ref{eq1.18}) and in~(\ref{xi0}),  the complete integration in space $\rho_k$ is
performed in the partition function $\Xi$, with the exception of integration over the
variable $\rho_0$ in $\Xi_{\rho_0}$.

\subsection{The generalized chemical potential $\mu^*$}

Our task here is to study the integral over $\rho_0$ in the expression~(\ref{xi0}).
After substitution $\rho_0 = \sqrt{N}\rho'_0$, omitting the terms
proportional to $\ln N$,
\be
\label{eq2.12}
\int \exp \left[N \left( \mu^* \rho'_0 + D \rho_0^{2'} - G\rho_0^{4'} \right)\right] \rd\rho'_0 = \int \exp [NE (\rho_0)] \rd\rho_0\,,
\ee
where
\be
\label{eq2.13}
E(\rho_0) = \mu^*\rho_0 + D\rho_0^2 - G\rho_0^4\,.
\ee

 Hereafter, we omit the primes.

Values $D$ and $G$ are now specified and they turn out to be positive. Coefficient at
$\rho_0^2$ in~(\ref{eq2.13}) is positive and the integrand increases at small
$\rho_0$, whereas at $\rho_0 \to \infty$ the function $\exp \{N
E_0(\rho_0)\}$ tends to zero due to the term $G\rho_0^4$.

Integral~(\ref{eq2.12}) is a function of the generalized chemical potential $\mu^*$, density $\eta$ and temperature $\tau$.

In the thermodynamical limit $N \to \infty$, $V \to \infty$, $\frac{N}{V} = {\rm const}$, the maxima of the integrand in~(\ref{eq2.12}) are very high. Therefore, the integral should be calculated by the steepest-descent method.

To do this, first we find the maximum of $E(\rho_0)$:
\bea
\label{eq2.14}
&& \frac{\partial E}{\partial \rho_0} = 0; \qquad  \frac{\partial^2 E}{\partial \rho_0^2} < 0\quad {\mbox or}\quad   \mu^* + 2 D \rho_0 - 4G \rho_0^3 = 0, \qquad  2D - 12G\rho_0^2|_{\rho_0 = \rho_{0\,\mathrm{max}}} < 0.
\eea

Thus, we have got an important result, the value for $\mu^*$:
\be
\label{eq2.15}
\mu^* = \left(- 2D \rho_0 + 4 G \rho_0^3\right)_{\rho_0 = \rho_0^{\mathrm{max}}}\,,
\ee
where for $\rho_0$ we have to take its value in the point of the absolute maximum of the integrand in~(\ref{eq2.12}).

Continuing our consideration, we shall write the equation~(\ref{eq2.14}) in a standard form:
\be
\label{eq2.16}
\rho_0^3 + V\rho_0 + W = 0.
\ee
Here,
\[
V = \frac{-D}{2G}\,, \qquad  W = -\frac14 \frac{\mu^*}{G}
\]
and $3\rho_0^2 + V > 0$, which corresponds to~(\ref{eq2.14}). Equation~(\ref{eq2.16}) has three solutions, that
may be found by Cardano formula:
\be
\label{eq2.17}
\rho_0 = \sqrt[3]{-\frac{W}{2} + \sqrt{Q}} +
\sqrt[3]{-\frac{W}{2} - \sqrt{Q}} ~,
\ee
$Q$ is a discriminant of the equation:
\be
\label{eq2.18}
Q = \frac{W^2}{4} + \frac{V^3}{27}\,.
\ee
The first term in the discriminant is always positive, the second one is always negative.
Thus, three possibilities may be observed: $Q > 0$, $Q < 0$ and $Q =
0$. Depending on the sign of $Q$, we have one real ($Q > 0$) or
three real solutions ($Q < 0$) \,\footnote{At $T>T_{\mathrm{c}}$ the discriminant
$Q$ is always positive, $Q > 0$.}.

Let us start with the limiting case $Q=0.$ This equality describes the intermediate surface between two regions  $Q>0$ and $Q<0$. Equation~(\ref{eq2.16}) has three real roots:
\be
\label{eq2.19}
\rho_1 = u + v, \qquad  \rho_2 = \rho_3 = - \frac12 (u+v),
\ee
where $u = \left[ - \frac{W}{2} + \sqrt{Q} \right]^{1/3}$, $v = \left[ - \frac{W}{2} - \sqrt{Q} \right]^{1/3}$. But only for the root $\rho_1$ we get a maximum for $E(\rho)$
\[
E''(\rho_1) = - GD < 0, \qquad  E''(\rho_2) = E''(\rho_3) = 0.
\]
Therefore, we take for $\rho_0 = \rho_0^{\mathrm{max}}$ in~(\ref{eq2.12})
\be
\label{eq2.20}
\rho_0^{\mathrm{max}} = \rho_1 = 2 \sqrt[3]{- \frac{W}{2}} = \sqrt[3]{\frac{\mu^*}{G}}\,.
\ee
Written explicitly, the condition $Q=0$ takes on the form:
\be
\label{eq2.21}
\left( \frac{W}{2} \right)^2 = - \left( \frac{V}{3} \right)^3 \quad {\mbox {and}}\quad
\mu^* = \pm a = \pm G \left( \frac23 \frac{D}{G} \right)^{3/2}.
\ee

So, when the discriminant $Q$ equals zero, we receive the value of generalized chemical potential of the system.
 From equation~(\ref{eq2.17})
\be
\label{eq2.22}
\rho_1 = \sqrt[3]{- \frac{\pm a}{G}}=\pm b, \qquad b=\sqrt{\frac{2D}{3G}}\,,
\ee
and, consequently, for $\rho_1$ we have two values $\rho_1 = b$ for $\mu^* = a$, and $\rho_1 = -b$ for $\mu^* = -a$.

\begin{figure}[ht]
\centerline{\includegraphics[width=0.45\textwidth,angle=0]{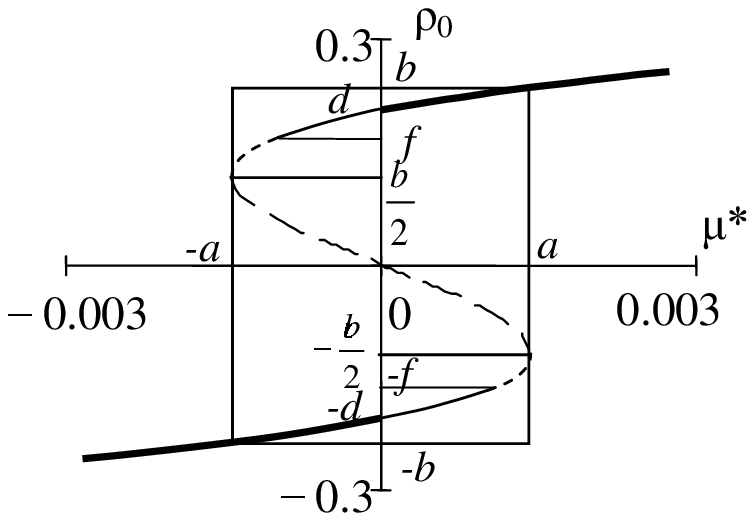}}
\caption{ \label{mu} The generalized chemical potential isotherm as a function of $\rho_0^{\mathrm{max}}$. Here,
$a = G \sqrt{\frac23 \frac{D}{G}}^{3} \sim \tau^{(5/2)\nu}$, $b =
\sqrt{\frac23 \frac{D}{G}} \sim \tau^{\nu/2}$, $d = \sqrt{\frac12 \frac{D}{G}} \sim \tau^{\nu/2}$, $f = \sqrt{\frac13 \frac{D}{G}}\sim\tau^{\nu/2}$.
}
\end{figure}

From the condition $Q=0$ we also receive:
\[
\rho_1 = b_0 |\tau|^{\nu/2}, \qquad b_0 = \sqrt{\frac23 \frac{D_0}{G_0}}
\]
and
\[
\mu^* = \pm \mu_0^* |\tau|^{5/2\nu}, \qquad  \mu_0^* = G_0 \left( \frac23 \frac{D_0}{G_0}\right)^{3/2}\,.
\]
Here we have two mutually reciprocal parabolic cylinder surfaces. The intersection with the plain $\tau = {\rm const}$ is a rectangle with the vertices $(-a,b)$, $(-a,-b)$, $(a,-b)$, $(a,b)$ as it is shown in figure~\ref{mu}.

For $\rho_2$ and $\rho_3$ we have $\rho_2 = - b/2$, $\rho_3 = b/2$, and $E''(\rho_2) = 12Gb$; $E'''(\rho_3) = -12Gb$.

In the case $Q>0$, equation~(\ref{eq2.15}) has one real and two complex solutions.
$Q>0$ means that
\[
\frac{W^2}{4} > \frac{V^3}{27} ~~~ {\mbox {\rm and}} ~~~ \frac{W}{2} > \sqrt{Q}.
\]
Thus,
\bea
&& \rho_0^{(1)} = \sqrt[3]{- \frac{W}{2}} \left\{ \left[ 1 - \frac{2\sqrt{Q}}{W} \right]^{1/3} +
\left[ 1 + \frac{2\sqrt{Q}}{W} \right]^{1/3} \right\}\,
. \nonumber
\eea
Expanding in powers of $\gamma$, where $\gamma = \left( \frac{V}{3} \right)^3 \Big/ \left( \frac{W}{2}
\right)^2, |\gamma| < 1$ we
receive:
\be
\label{eq2.23}
\rho_0^{(1)} = \left( \frac{\mu^*}{4G} \right)^{1/3} \left[ 1 + \left( \frac{|\gamma|}{4}\right)^{1/3} - \frac{1}{12} |\gamma| + \frac{\sqrt[3]{2}}{24} |\gamma|^{4/3} - 0 \left(\gamma^2\right) \right],
\ee
\[
\mu^* = q'G \Bigl(\rho_0^{(1)}\Bigr)^3, \qquad q' = 4 \left[ 1 + \left(  \frac{|\gamma|}{4} \right)^{1/3} - \frac{1}{12} |\gamma| + \dots \right]^{-3}.
\]
The sign of  $\rho_0^{(1)}$ is determined by the sign of $\mu^*$.
At  $|\gamma| = 1$, $Q=0$, $\rho_0^{(1)} = \sqrt[3]{\mu^*/G}$, $\rho_0^{(1)} = \pm b$ because
\be
\left( \frac14 \right)^{1/3} \left[ 1 + \left( \frac14 \right)^{1/3} - \frac{1}{12} +
\frac{\sqrt[3]{2}}{24} - \dots \right] = 1. \nonumber
\ee
So, at $Q=0$, $|\gamma| = 1$, the root $\rho_0^{(1)}$ coincides with the root $\rho_1$ from~(\ref{eq2.22}).

Thus, at $Q > 0$, both $\mu^*$ and $\rho_0^{(1)}$ vary within
\bea
\label{eq2.24}
&& |\mu^*| \geqslant  a, ~~ \mbox{where}~~ a = G \left( \frac23 \frac{D}{G} \right)^{3/2}, \nonumber\\
&& |\rho_0^{(1)}| \geqslant  b, ~~ \mbox{where}~~ b = \left( \frac23 \frac{D}{G} \right)^{1/2}.
\eea
In such a way, for $\rho_0^{(1)}$ we have two branches one for $\rho_0^{(1)} \leqslant  -b$ and negative values for $\mu^* \leqslant  -a$ and the second for $\rho_0^{(1)} \geqslant  b$ and $\mu^* \geqslant  a$.

 In  the region $Q < 0$, the equation~(\ref{eq2.14}) has three real solutions. It is more convenient to write them in trigonometrical form:
\bea
\label{eq2.25}
&&\mu^* = a \cos \varphi, \qquad \rho_{01} = b \cos \frac{\varphi}{3}, \qquad \rho_{02} = b \cos \frac{\varphi + 2\pi}{3}\,, \qquad \rho_{03} = b \cos \frac{\varphi + 4\pi}{3}\,, \nonumber \\
&&\varphi = \arccos t, \qquad t = - \frac{W}{2 \left( - \frac{V}{3} \right)^{3/2}} = \frac{\mu^*}{a}\,.
\eea

In the vicinity of the point $\mu^* = -a$ we have $\cos \varphi = -1$,
$\varphi = \pi$. Substituting the values $\varphi = \pi$ into the
solution, we obtain
 \bea
\rho_{01} = b\cos{\frac{\pi}{3}} = \frac{b}{2}\,, \qquad
\rho_{02} = b \cos \frac{3\pi}{3} = -b, \qquad
 \rho_{03} = b \cos \frac{5\pi}{3} = \frac{b}{2}\,. \nonumber
\eea
As we see, only the solution $\rho_{02}$ coincides with the solution
$\rho_{0}^{(1)}$ at the point $(-a-b)$.

In the vicinity $\mu^* = 0$, $\varphi = \frac{\pi}{2}$ and $\rho_{02} = b\cos \Bigl(\pi - \frac{\pi}{6} \Bigr) = - \frac{\sqrt{3}}{2}b$. Thus, the generalized chemical potential $\mu^*$ and the solution $\rho_0^{\mathrm{max}} = \rho_{02}$ varies within the intervals:
\[
-a \leqslant  \mu^* \leqslant  0, \qquad -b \leqslant  \rho_0^{(2)} \leqslant  -d \quad \mathrm{and} \quad  \pi \geqslant  \varphi \geqslant  \frac{\pi}{2}\,,
\]
where we denote $d = \frac{\sqrt{3}}{2}b = \sqrt{\frac12 \frac{D}{G}}$. At the point $(-a,-b)$ on the plane $\Bigl( \mu^* \rho_0^{\mathrm{max}}\Bigr)$ solutions $\rho_0^{(1)}$ and $\rho_{02}$ smoothly flow together.

For the case $\mu^* = a$, $\cos \varphi = 1$, $\varphi=0$ we have
\bea
 \rho_{01} = b \cos 0 = b, \qquad
 \rho_{02} = b \cos \frac{2\pi}{3} = -\frac12 b, \qquad
 \rho_{03} = b \cos \frac{4\pi}{3} = b \cos \left( \pi + \frac{\pi}{3} \right) = - \frac12 b. \nonumber
\eea

Now we have to take the solution $\rho_{01}$. It coincides with the solution $\rho_0^{(1)}$ at the point $\mu^* =a$, $\rho_{01} =b$, $\varphi =0$. At the point $\mu^* =0$ $\varphi = \frac{\pi}{2}$, the solution $\rho_{01}$ is $\rho_{01} = b \cos \frac{\pi}{6} = \frac{\sqrt{3}}{2}b$, or $\rho_{01} = d$.
So, in the interval $0 \leqslant  \mu^* \leqslant  a$, the solution $\rho_{01}$ varies inside the interval $d \leqslant  \rho_{01} \leqslant  b.$

We have got some significant result: on the axis $\mu^* =0$, the solution $\rho_0^{\mathrm{max}}$ varies jumping from the value $\rho_0^{\mathrm{max}} = \rho_{02} = -d$ to the value $\rho_0^{\mathrm{max}} = \rho_{01} = d$.

The plot of the generalized chemical potential isotherm as a function of $\rho_0^{\mathrm{max}}$
has the form shown in figure~\ref{mu}. Here, we have a smooth continuation of curves
$\rho_0^{(1)}$ to $\rho_{02}$ at the point $(-a,-b)$ and $\rho_0^{(1)}$ to $\rho_{01}$ at the point $(a,b).$

It is very important to note here that among the solutions~(\ref{eq2.25}), only solution $\rho_{02}$ in the region $-b < \rho_0 < -d$ and only solution $\rho_{01}$ in the region $d < \rho_0 < b$ obey conditions for an absolute  maximum of the function $E(\rho_0)$. For all other values of $\rho_{01}$, $\rho_{02}$, $\rho_{03}$, the absolute maximum of $E(\rho_0)$ in~(\ref{eq2.12}) cannot be realized.

Let us now find the slops of the generalized chemical potential isotherms $\mu^*(T, \rho_0^{\mathrm{max}})$ at the points $\mu^* = \mp a$, $\rho_0^{\mathrm{max}} = \mp b$,  and at the points $\mu^* = \mp 0$, $\rho_0^{\mathrm{max}} = \mp \frac{\sqrt{3}}{2}b = \mp d$. From~(\ref{eq2.14}) we have
\[
\frac{\rd\mu^*}{\partial \rho_0^{\mathrm{max}}} = - 2D + 12 G (\rho_0^{\mathrm{max}})^2\,.
\]

Thus, at the point $\mu^* = - a$, $\rho_0^{\mathrm{max}} = - b$, from the left, and
At the point $\mu^* = a$, $\rho_0^{\mathrm{max}} = b$ from the right,
\[
\frac{\rd\mu^*}{\partial \rho_0^{\mathrm{max}}} = 6D \sim \tau^{2\nu}
\]
and at the points $\mu^* = \mp 0$, $\rho_0^{\mathrm{max}} = \mp d$
\[
\frac{\rd\mu^*}{\partial \rho_0^{\mathrm{max}}} = 4D \sim \tau^{2\nu}.
\]

As we see, the slope of the tangents tend to zero proportionally to $\tau^{2\nu}$.
We have completed investigations of the generalized chemical potential $\mu^*$.

Having studied the integral~(\ref{eq2.12}), $\int \exp [N E (\rho_0)] \rd\rho_0$, and
the function $E(\rho_0)$, presented in~(\ref{eq2.13}), we revealed
the most essential changes in the behavior of the partition
function as well as in the behavior of thermodynamic functions.

In order to describe the scenario of the phase transition at $T \leqslant
T_{\mathrm{c}}$, we have to extract from the entire set of integration results
those that correspond to the integration the variables $\rho_0$ and $\omega_0$. This will automatically concern the events
connected with the generalized chemical potential $\mu^*$.

Our main goal in this study is to describe what exactly is happening at $T \leqslant
T_{\mathrm{c}}$. Here, in accordance with~\cite{21}, we restrict ourselves to the narrow region around the
critical point. The scenario of the phase transition is connected with
the behavior of the generalized chemical potential $\mu^*$.
 As it follows from equation~(\ref{xi0}), the plane $\mu^* = 0$
contains the coordinates of the critical point $T_{\mathrm{c}}, \eta_{\mathrm{c}}$.

\subsection{The partition function in the grand canonical ensemble at $T \leqslant  T_{\mathrm{c}}$}

Bringing together the obtained results, and taking the terms containing $\mu^*$,
 let us write the initial
partition function $\Xi$, according to equations~(\ref{xi_prod}), in
the form of a product of two partition functions:
\be
\label{eq2.26}
 \Xi = \Xi^{(2)} \Xi_{\rho_0}\,, \ee
where in $\Xi^{(2)}$ we included the results
\[
\Xi^{(2)} = \Xi_0 \Xi_{\mathrm{G}} \Xi_{\mathrm{L}}^{\rho_0}.
\]
Expression $\Xi^{(2)}$ does not
depend on $\mu^*$. All terms and effects connected with the behavior
of $\mu^*$, are gathered in the part $\Xi_{\rho_0}.$

\begin{figure}[ht]
\centerline{\includegraphics[width=0.65\textwidth,angle=0]{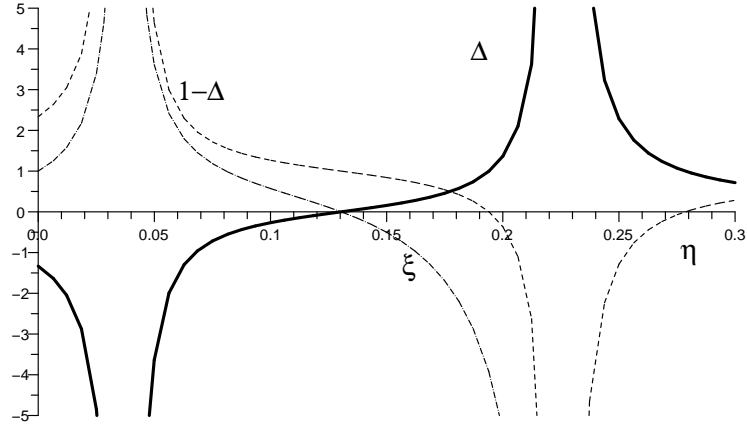} }
\caption{\label{delta} The dependence of $\Delta$ (solid line),$1-\Delta$ (dashed line), and $\xi$ (dash-dotted line) on the density $\eta.$ In the present work, the system is considered within the range $0.05<\eta <0.2.$ In this interval, the values of cumulant ${\frak{M}}_4$ are finite and negative.
}
\end{figure}

Partition function $\Xi_{\rho_0}$ is of most interest to us. From~(\ref{xi0})
\be
\label{eq2.27}
\Xi_{\rho_0} = \exp \Bigl[ \mu^* (1-\Delta)N \Bigr] \int
\re^{NE(\rho_0)}\rd\rho_0\,,
\ee
where
\[
 (1 - \Delta)N = \sqrt{N} \tilde {\frak M_1}\,, \qquad \Delta = - \Bigl( {\frak M}_2 \xi + \frac13 {\frak M}_3 \xi^2 \Bigr), \qquad E(\rho_0) = \mu^* \rho_0 + D \rho_0^2 - G \rho_0^4\,. \nonumber
\]
The dependence of $\Delta$ and $\xi$ on the density $\eta$ is shown in figure~\ref{delta}.

Our consideration is related with the thermodynamical limit $N \to \infty$, $V \to \infty$, $\frac{N}{V} = {\rm const}$. As a consequence, in the integral~(\ref{eq2.27}) we have to use the steepest-descent method, and regard only the point of the absolute maximum of the function $\exp \left\{ NE (\rho_0) \right\}$. Then,
\be
\label{eq2.28}
\left. \int\limits_{-\infty}^{\infty} \exp \Bigl\{ NE (\rho_0) \Bigr\}  \rd\rho_0 =
\exp \Bigl\{ NE (\rho_0) \Bigr\} \right|_{\rho_0 = \rho_0^{\mathrm{max}}}.
\ee

In the thermodynamical limit, the value of $\rho_0^{\mathrm{max}}$ coincides with the average value of $\rho_0$:
\bea
&& \langle \rho_0 \rangle = \frac{\int \rho_0 \exp \Bigl\{ NE (\rho_0) \Bigr\} \rd\rho_0}{\int \exp \Bigl\{ NE (\rho_0) \Bigr\} \rd\rho_0} = \frac{1}{N} \frac{\partial}{\partial \mu^*} \ln \int \exp \Bigl\{ NE (\rho_0) \Bigr\} \rd\rho_0\,, \nonumber\\[2ex]
&& \lim\limits_{N \to \infty} \langle \rho_0 \rangle = \rho_0^{\mathrm{max}}. \nonumber
\eea

\begin{figure}[!h]
\centerline{\includegraphics[angle=0,width=0.95\textwidth]{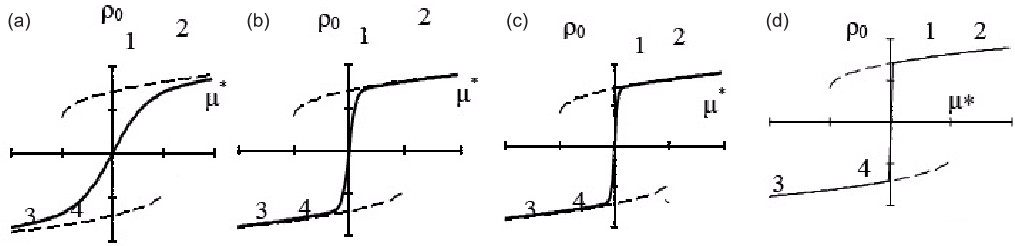}}
\caption{The average (solid line) $\langle \rho_0 \rangle$ and the most probable (dashed line) $\rho_0^{\mathrm{max}}$ values of $\rho_0$ at $N \to \infty$. The arcs 1--2 and 3--4 (dashed lines) correspond to absolute maxima of $\exp [NE(\rho_0)]$. Here, (a) $N=1000$; (b) $N=5000$; (c) $N=10000$; d) $N=100000$.
} \label{rho0}
\end{figure}

Illustration for this at $\tau = - 10^{-3}$ for argon is shown in figure~\ref{rho0}.

Hence, for expression~(\ref{eq2.27}) we can write:
\be
\label{eq2.29}
\Xi_{\rho_0} = \exp \left\{ N\left[\mu^* \left(1-\Delta\right) + \mu^*\rho_0^{\mathrm{max}} + D\left(\rho_0^{\mathrm{max}}\right)^2 - G\left(\rho_0^{\mathrm{max}}\right)^4\right] \right\},
\ee
where
\bea
\label{eq2.30}
&&  \rho_0^{\mathrm{max}} = \rho_1 \quad  \mbox{from}\quad (\ref{eq2.20}), \quad \mbox{when}\quad Q=0 \quad \mbox{and}\quad |\rho_1| =b, \qquad |\mu^*| = a, \nonumber\\
&&  \rho_0^{\mathrm{max}} = \rho_0^{(1)} \quad \mbox{from}\quad (\ref{eq2.23}), \quad\mbox{when}\quad Q>0 \quad\mbox{and}\quad |\rho_0^{(1)}| >b,\qquad |\mu^*| > a,
\nonumber\\
&& \rho_0^{\mathrm{max}} = \rho_{02}, \quad  -b \leqslant  \rho_{02} \leqslant  -d, \quad  \mu^* = a\cos \varphi, \quad  \frac{\pi}{2} \leqslant  \varphi \leqslant  \pi \quad  \mbox{from} \quad (\ref{eq2.25}), \nonumber\\
&& \rho_0^{\mathrm{max}} = \rho_{01}, \quad  d<\rho_{01} < b, \quad  \mu^* = a\cos \varphi, \quad  0 \leqslant  \varphi \leqslant  \frac{\pi}{2} \quad  \mbox{from} \quad (\ref{eq2.25}),
\eea
here, $ Q<0$. In all cases $\mu^* = - 2D \rho_0^{\mathrm{max}} + 4G (\rho_0^{\mathrm{max}})^3$.

Now we need to know the explicit values of $\rho_0^{\mathrm{max}}$ as function of density $\eta$. We use the condition
\be
\label{eq2.31}
\frac{\partial \ln \Xi}{\partial \beta \mu} = \frac{\partial \ln \Xi_{\rho_0}}{\partial \mu^*} = N.
\ee

From~(\ref{eq2.26}), (\ref{eq2.27}) and~(\ref{eq2.28})  we have
\be
N\left(1-\Delta + \rho_0^{\mathrm{max}}\right) \Bigr|_{\rho_0 = \rho_0^{\mathrm{max}}} = N. \nonumber
\ee
Hence,
\be
\label{eq2.32}
\rho_0^{\mathrm{max}} = \Delta.
\ee
Here, $\Delta = - \left( {\frak M_2}\xi + \frac13 {\frak M_3}\xi^2 \right)$, $\xi = \frac{{\frak M_3}}{|{\frak M_4}|}$; cumulants ${\frak M_2}$, ${\frak M_3}$, ${\frak M_4}$ are the known functions of a density (see table~\ref{cum_tab}).

Finally, we obtain
\bea
\label{eq2.33}
\Xi_{\rho_0} = \exp \left\{ N\left[\mu^* + D\Delta^2 - G\Delta^4\right] \right\}
\eea
and
\be
\label{eq2.34}
\mu^* = - 2D\Delta + 4G(\Delta)^3.
\ee

All components in expression~(\ref{eq2.26}) and~(\ref{eq2.27}) are defined.
Thus, all principal problems in calculating the partition function of the system with Lennard-Jones type interaction are solved.

\section{The main results}

Now we are going to determine the expression for the critical point, the critical region on the whole, the events inside the region $-d \leqslant  \Delta \leqslant  d$ where the boiling process occurs, the notion about the overcooled gas and overheated liquid, and the equality of the chemical potentials at $\Delta = \pm d$, and finally the comparison with the experimental data for some substances.

\paragraph{The rectilinear diameter.} The expression $\Delta(\eta) =0$ gives us a starting point for the rectilinear diameter. The equation $\Delta =0$ reads
\be
\label{eq2.35}
{\frak M_2}(\eta) \frac{{\frak M_3}(\eta)}{|{\frak M_4}(\eta)|} - \frac13 {\frak M_3}(\eta)
\left( \frac{{\frak M_3}(\eta)}{{\frak M_4}(\eta)} \right)^2 =0
\ee
and $\mu^*(\Delta) = -2D\Delta + 4G\Delta^3 =0$.

In the initial expression~(\ref{eq2.8}) for the partition function, this equation means that we consider the Ising-like case and that we are looking for the critical point of the phase transition of the second order. Taking into account formulas for ${\frak M_n}(0),$ the plots for cumulants in figure~\ref{cumulant} and the dates from table~\ref{cum_tab}, we get from~(\ref{eq2.35}) some universal quantity for the critical density
\be
\label{eq2.36}
\eta_{\mathrm{c}} = 0.13044.
\ee
This is the very localization of the rectilinear diameter on the axis $\eta$.

{\it The critical temperature} expression follows from the whole solution to our problem. At the critical temperature integration of the partition function $\Xi_{\mathrm{L}}$ in~(\ref{eq2.8}) is connected with the existence of the Dyson's hierarchical symmetry between the block Hamiltonians, and with making use of the Wilson's linearization method for solving the recursion equations. The integration in~(\ref{eq2.8}) should be fulfilled in the system of the block Hamiltonians. At the critical point, it is made completely in the critical regime for the whole interval $0 \leqslant  k \leqslant  B$. For the sequence of coefficients $d^{(n)}(k)$ and $a_4^{(n)}$ at the critical point, we have situation pictured in figure~\ref{integration}~(a). It is scrupulously described in~\cite{8,8a,9,9a} and we shall take the ready made formula for $T_{\mathrm{c}}$ therefrom
\be
\label{eq2.37}
T_{\mathrm{c}} (\eta) = \frac{|\bar \alpha(0)|}{k_{\rm B}} \cdot  \frac{2\left[ 1 - \bar r + R_{12}^0 \sqrt{\bar u}\big/(R_{11} - E_2) \right] }
{a_2 + \left\{ a_2^2 + \left[ 4a_4 R_{12}^0 \sqrt{\bar u}\big/(R_{11} - E_2) \right] \left[ 1 - \bar r + R_{12}^0 \sqrt{\bar u}\big/(R_{11} - E_2)\right] \right\}  } \,,
\ee
$\alpha(0) = \beta_{\mathrm{c}} \bar \alpha(0)$, $a_2$ and $a_4$ are the initial coefficients from the expression~(\ref{eq2.8}), all other values are given in table~\ref{coefs}.
\begin{table}[h]
\caption{Values of some coefficients of the solution of recursion relations for $s=3.58$.}
\label{coefs}
\begin{center}
\begin{tabular}{|c|c|c|c|c|c|c|c|}
\hline
$E_1$ & $E_2$ & $\bar{r}$ & $\bar{u}$ & $R_{12}^{(0)}$ & $R_{21}^{(0)}$ & $q$   & $R_{11}$ \\
\hline\hline
$8.235$ & 0.377 & 0.612 & 0.889 & 3.837 & 1.174 & 0.612 & 7.613 \\
\hline
\end{tabular}
\end{center}
\end{table}

\begin{figure}[ht]
\centerline{\includegraphics[angle=0,width=0.45\textwidth]{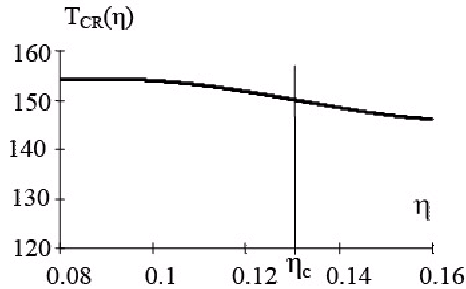}}
\caption{Coordinates of the liquid-gas critical point on the $\mu^*=0$ plane for argon.
} \label{CP}
\end{figure}
This formula is valid for the Ising model with the fixed initial coefficients $a_2$, $a_4$, $\alpha(0)$. In our case, they are functions of density $\frac{\langle N \rangle}{V}$. Therefore, formula~(\ref{eq2.37}) describes the surface of critical temperatures. Its intersection with the plane $\mu^* =0$ gives us the curve of critical temperatures and the intersection with the rectilinear diameter $\Delta = 0$ or $\eta_{\mathrm{c}} = 0.130443$ gives us the critical point, as it is demonstrated in figure~\ref{CP}. In the case of argon equation~(\ref{eq2.35}), conditions $\mu^* =0$, and $\Delta =0$ gives us:
\[
\eta_{\mathrm{c}} = 0.130443,\qquad \frac{k_{\rm B} T_{\mathrm{c}}}{\varepsilon} = 1.25, \qquad T_{\mathrm{c}} = 123.27~\mathrm{K}.
\]
Finishing the consideration about the critical point $T = T_{\mathrm{c}}$, $\eta = \eta_{\mathrm{c}}$ we shall gather together all three conditions determining the critical point
\bea
\label{eq2.38}
&& T_{\mathrm{c}} = \frac{|\bar \alpha(0)|}{k_{\rm B}} \frac{2R_1}{a_2 + \{ a_2^2 + R_2R_1 \}}\,, \nonumber\\
&& \mu^* = \beta (\mu - \mu_0) - \xi + |\alpha(0)|(1 - \Delta) = 0, \nonumber\\
&& \Delta = - \Bigl( {\frak M_2}\xi - \frac13 {\frak M_3}\xi^2 \Bigr) = 0.
\eea

We note here
\[
R_1 = 1 - \bar r + R_{12}^0 \sqrt{\bar u}(R_{11} - E_2)^{-1}, \qquad R_2 = 4a_4 R_{12}^0 \Bigl[ \sqrt{\bar u}(R_{11} - E_2) \Bigr]^{-1}.
\]

In table~\ref{tab_a} we present the results of calculation of the effective hard sphere diameter and the values of the
critical temperatures. The values of the coefficient $a_2$ and $a_4$ from table~\ref{cum_tab} are taken into account. As may be seen,  the accordance with the experiments data is quite satisfactory.

\paragraph{Critical region.} At the critical point $\tau =0$, $\eta = \eta_{\mathrm{c}}$, the linear approximation for the recursion equations, using the fixed point as a particular solution to them, is valid in the whole region of $k,$ $0 \leqslant  k \leqslant  B$. When $\tau$ is different from zero, the linear approximation is valid only for some interval $k$, $B_{m_\tau} < k < B$, where $B_{m_\tau} = B/s^{m_\tau}$. This means that only in this interval of $k$ the renormalization-group type solutions are valid. Only these solutions reflect the renormalization-group symmetry. Here we refer to the interval $B_{m_\tau} \leqslant  k \leqslant  B$ as a critical-regime-interval, and the cyclic semigroup symmetry between coefficients $d^{(n)}(B_{n+1}B_n)$, $d^{(n-1)}(B_n B_{n-1})$ and $a_4^{(n)}$, $a_4^{(n-1)}$ as a critical regime (CR). In [9,14], there were found the quantities $m_\tau$ determining $B_{m_\tau}$ and the interval of temperatures $\tau$, $\tau^* \geqslant  \tau \geqslant  0$ and $\tau^* \leqslant  \tau \leqslant  0$, containing the CR.

For $\tau < 0$ we have~\cite{9}
\be
\label{eq2.39}
(\tau^*)^{\nu} \left( \frac{\tilde C_1}{r^* + \alpha(0)} \right)^{\nu} = \sqrt{1 - \frac{a_2}{\alpha(0)}}
\ee
for the $s = s^*$, $\tau^* \simeq 0.02$. In the l.h.s. of (\ref{eq2.39}), there is a quantity inversely proportional to the correlation radius in the critical regime. In the r.h.s. of (\ref{eq2.39}), there is a quantity inversely proportional to the correlation radius  in the limit Gaussian regime. At $n=m_{\tau}$ the both of them are equal. At $k<B/s^{m_{\tau}},$ after performing the shift $\rho_0=\rho'_0+\sigma$ in expression for $\Xi_{\mathrm{L}},$ the Gaussian density measure is the basic one, and at $B/s^{m_\tau}\leqslant  k \leqslant  B$ the fourfold density measure is the basic measure.

Partition function consists of two parts, one belonging to the CR and another one to the inverse-Gaussian-regime. From the equation~(\ref{eq2.38}) we can get the temperature boundary of the CR. The density-size of the critical regime at $T < T_{\mathrm{c}}$ is closely linked with the temperature $\tau^*$. The magnitude of $\eta^*$  as the boundary of density of the CR may be taken from the expression:
\be
\label{eq2.40}
\Delta (\eta^*) = b(\tau^*)
\ee
or in the explicit form:
\[
- \left[ {\frak M_2} \frac{{\frak M_3}}{|{\frak M_4}|} + \frac13 {\frak M_3}
\left( \frac{{\frak M_3}}{{\frak M_4}} \right)^2 \right]_{\eta = \eta^* = \eta(\tau^*)} =
\sqrt{\frac23 \frac{D_0}{G_0}}(\tau^*)^{\nu/2}
\]
at $\tau =0$, $\eta = \eta_{\mathrm{c}} = 0.13044$. $\eta^* = \eta(\tau^*)$ is the solution of the equation~(\ref{eq2.35}).

\begin{figure}[h]
\centerline{\includegraphics[angle=0,width=0.7\textwidth]{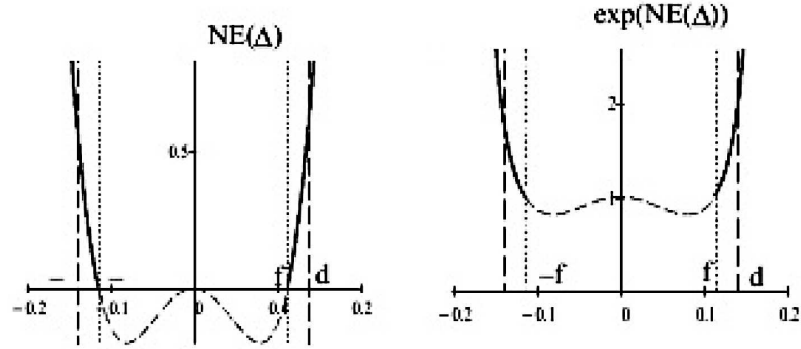}}
\caption{The curves of the function $NE(\Delta)$ in the region $-d \leqslant  \Delta \leqslant  d$. The most important points: $\Delta = \pm d$ the jump-like transition gas-liquid or liquid-gas; $\Delta = \pm f$ the points for the extreme overcooled gas or extreme overheated liquid. The region $-f < \Delta < f$ is the unattainable region of densities, $\Delta =0$ is localization of the rectangular diameter. The order parameter is equal to $2d = \sqrt{2D/G} \sim \tau^{\nu/2}$, $N=2000.$
} \label{expon}
\end{figure}

\paragraph{The overheated liquid and the overcooled gas, the order parameter.} We need to describe more in detail the function $\Xi_{\rho_0}(\Delta)$ in formula~(\ref{eq2.33}) for $\Delta$ the interval $-d \leqslant  \Delta \leqslant  d$.
\[
\Xi_{\rho_0} = \exp [\mu^*(1-\Delta)] \int \exp \left[NE (\rho_0)\right] \rd\rho_0\,.
\]
Here, $E (\rho_0) = \mu^* \rho_0 + D\rho_0^2 - G\rho_0^4, ~~ \rho_0^{\mathrm{max}} = \Delta$ and $\mu^* = -2D\Delta + 4G\Delta^3$ for $\rho_0 = \rho_0^{\mathrm{max}}$. To calculate the integral $\int \exp [NE (\rho_0)] \rd\rho_0$ we can use the steepest-descent method only when $E(\Delta)=\mu^*\Delta + D\Delta^2-G\Delta^4$ is positive. To this end, we shall look for the points where $E(\Delta) =0$:
\[
(-2D\Delta + 4G\Delta^3)\Delta + D\Delta^2 - G\Delta^4 = 0.
\]
We get three points $\Delta =0$, $\Delta = \pm f$, where $f = \sqrt{D/3G}$. The curves for both $E(\Delta)$ and $\exp[NE(\Delta)]$ are demonstrated in figure~\ref{expon}.

Let us discuss the situation around figure~\ref{expon}. At the points $\Delta = \pm b$ [discriminant $Q=0$, see~(\ref{xi_prod})--(\ref{eq2.17})] we have a boundary between the single states of homogeneous gas (for $\Delta < -b$) and of homogeneous liquid (for $\Delta >b$) and a ``process'' of two phases arising.

At the points $-d \leqslant  \Delta \leqslant  d$, $\mu^* =0$, there occurs a jump of densities, i.e., the process of boiling. This is a situation of emitting (or consuming) the latent heat.

The probabilities of the state when $|\Delta| = b$ and $|\Delta| = d$, are proportional to $\exp [NE(b) ]= \exp \left( N \frac23 \frac{D^2}{G} \right)$ and $\exp [NE(d)] = \exp \left( N \frac14 \frac{D^2}{G} \right)$. There are the points of high probability. The points $\Delta = \pm f$ are the least attainable points, the $\exp [NE(\Delta)] =1$ for $\Delta =0, \pm f$.

The states for which $\Delta = \pm b/2$ have $E \Bigl(\frac{b}{2} \Bigr) = - \frac{1}{12} \frac{D^2}{G}$, and the probability $\exp \Bigl[ -N\frac{D^2}{12G}\Bigr]$ tend to zero when $N \to \infty$. The states at these points are absolutely not attainable.

So, we can say that the states for which  $-d < \Delta < -f$, are the states of overcooled gas, and the states for which $f < \Delta <d$ are the  states of overheated liquid.

\paragraph{The order parameter.} All significant points $b, d$ and $f$ are connected with the phase transition processes. This really takes place in our system. And there is a question: what is the {\it order parameter}. It may be $\Delta = b$, or $\Delta =d$. The quantity $\Delta_{\eta_0}^{(d)} = \eta_{\mathrm{L}}^{(d)} - \eta_{\mathrm{G}}^{(d)} = 2d(\eta)$, where $d = \sqrt{\frac12 \frac{D}{G}}$ represents the difference of the density between liquid and gas states at the beginning and at the end of the boiling process.

The quantity $\Delta \eta^{(b)} = \eta_{\mathrm{L}}^{(b)} - \eta_{\mathrm{G}}^{(b)} = 2b(\eta)$ where $b = \sqrt{\frac23 \frac{D}{G}}$ is the difference between the liquid and the gas densities at the moment when the two phase gas-liquid situation arises and at the moment of its disappearance inside the monophasic gas or liquid states.

We take here the quantity
\[
2d = \sqrt{2D/G}
\]
as the order parameter of the system. Now we are going to talk about the equality of the chemical potentials at the beginning and at the end of the boiling process.

{\it The equality of chemical potentials. Equilibrium conditions}. Accor\-ding to~(\ref{eq2.5}), the generalized chemical potential $\mu^*$ is equal to
\[
\mu^* = h - \xi + |\alpha(0)| \frac{\tilde {\frak M_1}}{\sqrt{N}}\, ,
\]
where
\[
 h = \beta (\mu - \mu_0), \qquad  \xi = \frac{\frak M_3}{|\frak M_4|}\,, \qquad  \tilde {\frak M_1} = \sqrt{N} (1- \Delta), \qquad \alpha(0) = \frac{N}{V} \frac{\tilde \Phi(0)}{k_{\rm B} T}\,.
\]

\begin{figure}[ht]
\centerline{\includegraphics[angle=0,width=0.45\textwidth]{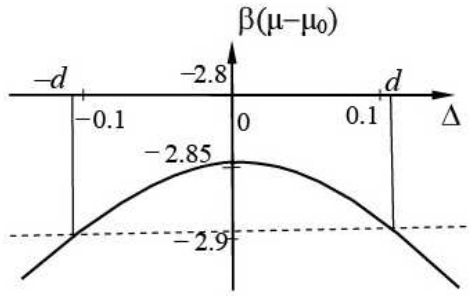}}
\caption{Chemical potential $\beta(\mu-\mu_0)$ on the binodal $\Delta=d$ for Argon at $|\tau|=0.001.$ Dashed line indicates two points on the binodal: $-d$ is transition to the gas state, $d$ is transition to the liquid state.} \label{binodal}
\end{figure}

Near the points of the phase transition of the first order,
$\Delta_{\mathrm{G,L}}=\mp \sqrt{\frac12 D/G}$, the function $\mu^*$ tends to zero, and
\be
\label{eq2.41}
\left\{\beta \mu =   \beta \mu_0 + \xi - |\alpha(0)| (1 - \Delta)\right\} \Big|_{\Delta_{\mathrm{G}} = - \sqrt{\frac12 D/G}, \,\,  \Delta_{\mathrm{L}} =  \sqrt{\frac12 D/G}}\,.
\ee
Here, all the functions on the right hand side are monotonous functions of density $\eta$, of the  parameter $\Delta$, and of temperature $\tau$, see figure~\ref{delta} and~\ref{binodal}.

\paragraph{Comparison with the experimental data.} To this end, we have to know at first the parameters of the full initial Lennard-Jones-potential, pictured in figure~\ref{potential}. It consists of the potential of hard core (point $b$) with the parameter $\sigma$ and of the attractive long-range potential $\phi(r)$ (point $c$, and $\sigma_0$). To find parameter $\sigma$ (of the reference system) we equate the critical density $\eta_{\mathrm{c}}$ from~(\ref{eq2.37}) to the experimentally determined critical density $\delta_{\mathrm{c}}$ for the concrete matter:
\[
 \eta_{\mathrm{c}} = \frac{N_{\mathrm{c}}}{V} \frac{\pi \sigma^3}{6}\,,\qquad \frac{N_{\mathrm{c}}}{V} = \eta_{\mathrm{c}} \frac{6}{\pi \sigma^3}\,, \qquad \delta_{\mathrm{c}} = \frac{N_{\mathrm{c}}m}{V} = \frac{N_{\mathrm{c}}}{V} \frac{M}{N_{\mathrm{A}}}~~ {\rm g/cm^3},
\]
$m$ is the mass of the particle, $m = \frac{M}{N_{\mathrm{A}}}$, $M$ is molecular weight, $N_{\mathrm{A}}$ is Avogadro-number.
 Then,
\be
 \delta_{\mathrm{c}} = \frac{6\eta_{\mathrm{c}}}{\pi \sigma_0^3} \frac{M}{N_{\mathrm{A}}} \nonumber
\ee
and in figure~\ref{potential} we have to put for $\sigma$ the quantity
\be
\label{eq2.42}
\sigma = \left( \frac{6\eta_{\mathrm{c}}}{\pi \delta_{\mathrm{c}}} \frac{M}{N_{\mathrm{A}}} \right)^{1/3}.
\ee

Using expression~(\ref{eq2.36}), we can make a comparison of theoretical results with the experimental data (for several substances, see table~\ref{tab_a}).

In figure~\ref{binodal} the isotherm of the chemical potential $\mu$ is given for argon. The curve is quite symmetrical with regard to the rectilinear diameter $\Delta(\eta) =0$, $\eta_{\mathrm{c}} = 0.13044$. The horizontal lines intersects the curve at the points of equal values of chemical potential $\mu$.

In such a way, we have used the experimental value for the critical density $\rho_{\mathrm{c}}$ to determine the constants of the initial potential in~(\ref{eq1.5}) and~(\ref{eq1.7}). We had to take into account equation~(\ref{eq2.35}) for the $\eta_{\mathrm{c}}$ and $\eta_{\mathrm{c}} = 0.13044$. The condition $\Delta =0$ automatically demands $\mu^* =0$ as it is seen from~(\ref{eq2.34}). So, the rectilinear diameter places on the surface $\mu^* =0$. The phase transition of the second order takes place on the surface. Here and always from~(\ref{eq2.35}) $\eta_{\mathrm{c}} = 0.13044$. For example for Argon, $M = 39.948$ $\delta_{\mathrm{c}} = 0.533$~g/cm$^3$ we get $\sigma = 3.2007$~\AA.

The attractive potential $\Phi(r)$ equals zero for all $r < \sigma_0$.

Expression~(\ref{eq2.35}) plays a critical role when doing comparison with the experiment (see table~\ref{tab_a}).

\begin{table}[h]
\caption{The critical temperature $T_{\mathrm{c}}$ and the effective hard sphere diameter $\sigma$ for some systems.}
\label{tab_a}
\begin{center}
\begin{tabular}{|c|c|c|c|c|c|c|}
\hline
System & $T_{\mathrm{c}},~{^\circ}$C & $T_{\mathrm{c}},~{^\circ}$C & $\sigma_0$, \AA & $\sigma$, \AA & $\sigma/\sigma_0$ & $\varepsilon/k_{\mathrm{B}}$, K \\
       &  (exp.)      & (this   & (exp.)          &   &  &   \\
       &             &  work)  &                 &   &  &   \\
\hline\hline
CO--CO       & --140.23 & --138.46 & 3.76 & 3.37 & 0.898 & 100.2  \\
Ar--Ar       & --122.65 & --123.27 & 3.405& 3.14 & 0.922 & 119.8  \\
Kr--Kr       & --63.1   & --67.84  & 3.6  & 3.367 &0.935 & 171    \\
Xe--Xe       & 16.62   & 16.84   & 4.1  & 3.71 & 0.905 & 221    \\
O$_2$--O$_2$ & --118.84 & --110.8  & 3.58 & 3.18 & 0.89  & 117.5  \\
N$_2$--N$_2$ & --147.05 & --150.02 & 3.698& 3.365 &0.91  & 95.05  \\
\hline
\end{tabular}
\end{center}
\end{table}


\section{Conclusions}

This work is a continuation of our previous articles printed in different journals~\cite{15,16,17,18,19,20,21}. In this article, we made some complete investigations of the behaviour of the liquid-gas system at the critical point $T = T_{\mathrm{c}}$ and below it, $T \leqslant  T_{\mathrm{c}}$. We worked in the narrow vicinity to the critical point in the region where the system obeys  a special symmetry-behaviour of the scale invariance referred to as the critical regime. We tried to describe the property of the gas-liquid systems and to compare our results with the experimental data for some real substances. To this end, for description we took the Lennard-Jones type potentials.

We made use of the collective variables method developed previously in solving the Ising model~\cite{7,8,8a,9,9a,14}. We worked in grand canonical ensemble. The repulsive part of the potential was taken into account including the reference system given on the Cartesian phase space of particles coordinates. The hard spheres system was used as the reference system. An estimation of the effective hard sphere diameter is proposed in~(\ref{eq2.42}).

The long-range attraction was described in the phase space of collective variables $\{ \rho_{\bf k}\}$. In such a way, the short-range and long-range interactions work in different phase-spaces. The natural ``crossing'' of the short-range and long-range interactions takes place in the equation $\frac{\partial \ln \Xi_{\mathrm{L}}}{\partial \mu^*} = N$ [see~(\ref{eq2.31}) and~(\ref{eq2.32})].

In the way the problem was stated, we were restricted to the region of
minimum of the Fourier-image of attraction potential (the wave
vectors $k \leqslant   B$). We assumed that the main events concerning the phase transition concentrate in this region. We supposed that the problem at $k > B$
had a known solution, which can be presented, e.g., in the form of the
virial series with convergent integrals. More accurate results could
have been produced by renormalization of the quantities
 $D$ and $G$ in the region of $k<B$. However, as was shown in~\cite{20}, the correction is inessential.

To describe the interaction between the particles at short-range
distances, the reference system of elastic particles is introduced
with the corresponding cumulant values $\frak M_1$, $\frak M_2$, $\frak
M_3$, $\frak M_4$. A principal question in solving the problem in
general is the discovery of wide plateaus  in cumulants $\frak M_n$
at small values of $k$ in the vicinity of the point $k = 0$~\cite{16}.
The interval (0B) of the wave-vectors ${\bf k}$ is located completely on the plate, and that of the cumulants ${\frak M_n}$ located at $k =0$.
Corresponding calculations permit us to reduce the problem of
gas-liquid critical point to the Ising model in external field~\cite{17,18}.

The effect of short-range interactions  concentrated in
a reference system is essential to solving our problem. The values $\frak
M_2(0)$, $\frak M_3(0)$ and $\frak M_4(0)$ produce an expression for
the parameter \linebreak
 $\Delta = - \left[ \frak M_2(0)\xi + \frac13 \frak M_3(0)\xi^2 \right]$, $\xi = \frac{\frak M_3(0)}{|\frak M_4(0)|}$, which is the main variable of the equation of state.  The natural ``crossing'' of the short-range and long-range interactions
takes place in the equation $\frac{\partial \ln \Xi_{\mathrm{L}}}{\partial \mu^*} =
N$ [see~(\ref{eq2.31}) and~(\ref{eq2.32})] under substitution
 of the generalized chemical potential $\mu^*$
by its values as a function of  $\tau$ and $\Delta(\eta)$.

Our analysis is correct in the region
of densities $\eta$, where the cumulant $\frak M_4$ is finite and
negative, namely at 0.02 $\leqslant   \eta \leqslant  $0.2.  Let us note that integration is
carried out in the phase space of collective variables $\rho_{\bf k}$,
$\rho_0$. That is why the values of initial  coefficients $a_2$ and
$a_4$, presented in table~\ref{cum_tab}, are essential.

In general, after a twenty-year long break, connected with the
political activities of one of us, we would like to express the heartfelt gratitude
to our friends, collaborators at the Institute for Condensed
Matter Physics of the NAS of Ukraine, in particular to I.M.~Mryglod and O.L.~Ivankiv,  for permanent assistance in  returning to the  ``liquid-gas critical point'' problem
 and for the first discussion of this work at the Institute seminar. We are sincerely grateful to L.A.~Bulavin for discussion of this work at the seminar of the physics faculty at T.~Shevchenko  Kyiv National university  and to A.G.~Zagorodny for the discussion of the work at the seminar of the Bogolubov Institute for Theoretical Physics NAS of Ukraine.

We are sincerely grateful to M.P.~Kozlovsky and especially to
R.~Romanik for fruitful discussions and for assistance
in proofreading the paper and for the preparation of some
illustrations, heartily thank  Yu.~Holovatch for useful discussions of
the results and for assistance in preparing an English version of
the paper, heartily thank O.V.~Patsahan for proofreading the paper
and for useful pieces of advice concerning calculations of latent
heat of transition.

As a result of this investigation a lot of different characteristics were obtained: the critical point, the critical temperature and the critical density; the order parameter; the sphere of the phase transitions; the size of the jump of the density during the phase transition of the first order; the size of the critical areas; the intervals of densities for the overcooled and overheated states; the equality of the chemical potentials at the beginning and at the end of the jump of density during the phase transition; the area of the phase transition of the second order near the critical point; the way to get the values of $\sigma$ -- the size of the diameter of the hard sphere of the particles, and finally the comparison with the experimental data for Ar, Kr, Ne, Xe, O$_2$, N$_2$, CO. We have got quite satisfactory results.

Of course, a lot of problems remain unresolved. We talk about the coordination of the mutual accuracy in the consideration of long-range and short-range interactions, about taking into account the dependence of cumulants ${\frak M_2}$ and ${\frak M_4}$ on $k^2$, about taking into account the attractive interactions in the area $k>B$ and other issues. A more precise consideration is to be undertaken.

\bibliography{References}

%
%

\ukrainianpart

\title{Фазовий перехід рідина-газ в критичній точці та в області нижче критичної точки}
\author{І.Р. Юхновський, В.О. Коломієць, І.М. Ідзик}
\address{
 Інститут фізики конденсованих систем НАН України, Львів, 79011, вул. Свєнціцького, 1
}

\makeukrtitle

\begin{abstract}
Ця стаття є продовженням наших попередніх робіт
(див. Yukhnovskii~I.R. et al., J.~Stat. Phys, 1995, \textbf{80}, 405, а також посилання там),
в яких  ми описали поведінку простої системи взаємодіючих частинок у критичній точці і в
області температур вище критичної точки, $T\geqslant  T_{\mathrm{c}}$.
Тут ми описуємо поведінку системи  в критичній точці $(T_{\mathrm{c}}, \eta_{\mathrm{c}})$
і в області температур нижче критичної точки.
Розрахунки здійснюються з перших принципів. Вираз для великої статистичної суми приведений до
функціонального інтегралу на множині колективних змінних і представлений в ізингоподібній формі.
Нижче   $T_{\mathrm{c}}$,  де система демонструє фазовий перехід першого роду, тобто кипіння,
відбувається ``стрибок'' між ``екстремально'' високими ймовірностями газового і рідкого станів, при цьому виділяється або поглинається  прихована теплоту переходу. Виведено також умови фазової рівноваги.

\keywords фазовий перехід рідина-газ, критична точка, колективні змінні

\end{abstract}

\end{document}